\documentclass[fleqn,usenatbib]{mnras}

\usepackage{newtxtext,newtxmath}

\usepackage[T1]{fontenc}
\usepackage{ae,aecompl}

\usepackage{graphicx}	
\usepackage{amsmath}	
\usepackage{amssymb}	
\usepackage[usenames, dvipsnames]{color} 

\title[Galaxy-galaxy lensing in CLASH]{Galaxy-galaxy lensing in the outskirts of CLASH clusters: constraints on local shear and testing mass-luminosity scaling relation}

\author[Desprez et al.]{Guillaume Desprez$^{1,2}$\thanks{E-mail: guillaume.desprez@unige.ch}, 
Johan Richard$^{2}$, 
Mathilde Jauzac$^{3,4,5}$,
Johany Martinez$^{2}$, 
\newauthor Brian Siana$^{6}$, 
Benjamin Cl\'ement$^{2}$ 
\\
$^{1}$Department of Astronomy, University of Geneva, ch. d'Ecogia 16, CH-1290 Versoix, Switzerland\\
$^{2}$Univ Lyon, Univ Lyon1, Ens de Lyon, CNRS, Centre de Recherche Astrophysique de Lyon UMR5574, F-69230, Saint-Genis-Laval, France\\
$^{3}$Centre for Extragalactic Astronomy, Department of Physics, Durham University, Durham DH1 3LE, U.K.\\
$^{4}$Institute for Computational Cosmology, Durham University, South Road, Durham DH1 3LE, U.K.\\
$^{5}$Astrophysics and Cosmology Research Unit, School of Mathematical Sciences, University of KwaZulu-Natal, Durban 4041, South Africa\\
$^{6}$Department of Physics \& Astronomy, University of California, Riverside, CA 92521, USA\\
}

\date{Accepted XXX. Received YYY; in original form ZZZ}

\pubyear{2015}

\begin{document}
\label{firstpage}
\pagerange{\pageref{firstpage}--\pageref{lastpage}}
\maketitle

\begin{abstract}
We present a selection of 24 candidate galaxy-galaxy lensing (GGLs) identified from Hubble images in the outskirts of the massive galaxy clusters from the CLASH survey. These GGLs provide insights into the mass distributions at larger scales than the strong lensing region in the cluster cores. We built parametric mass models for three of these GGLs showing simple lensing configurations, in order  to assess the properties of their lens and its environment. We show that the local shear estimated from the GGLs traces the gravitational potential of the clusters at 1-2 arcmin radial distance, allowing us to derive their velocity dispersion. We also find a good agreement between the strength of the shear measured at the GGL positions through strong-lensing modelling and the value derived independently from a weak-lensing analysis of the background sources. Overall, we show the advantages of using single GGL events in the outskirts of clusters to robustly constrain the local shear, even when only photometric redshift estimates are known for the source. We argue that the mass-luminosity scaling relation of cluster members can be tested by modelling the GGLs found around them, and show that the mass parameters can vary up to $\sim$~30\% between the cluster and GGL models assuming this scaling relation.

\end{abstract}




\section{Introduction}

Dark matter (DM) is one of the most challenging questions in modern astrophysics. It is indeed the most common matter species in the Universe according to the most commonly accepted model of cosmology, $\Lambda$CDM, but remains undetectable directly. Its abundance in the largest observable structures of the Universe such as galaxy clusters and massive galaxies, makes these systems ideal probes to understand its properties.

Galaxy clusters are the most massive collapsed objects observable, and their matter content is dominated by DM (up to $\sim$85\%). Due to their high mass, they will act as gravitational lenses, deflecting the light coming from galaxies located behind \citep[see][for some reviews]{Massey2010,Kneib2011,Hoekstra2013}. The geometry and location of these deflected images of background galaxies can be used to trace the dark matter distribution in these clusters.
In the core of clusters where the density is the highest, we observe highly magnified and multiple images of background galaxies, this is the strong-lensing regime \citep{Soucail1988a}. 
However, even for the most massive and concentrated cluster cores, the strong-lensing region remains small, up to $\sim 20-40$\arcsec\ (typically $<500$ kpc) from the cluster centre  (\citealt{Richard2010b,Zitrin2011,Merten2011,Richard2014,Jauzac2015a,Grillo2015}).
Extending outside this region, the density drops and the distortions are much smaller, this is the weak-lensing regime \citep[][]{Smith2005,Jauzac2012,Medezinski2013,Umetsu2015}. By combining both lensing regimes, we can trace the mass distribution of galaxy clusters up to a few Mpc radius \citep{Bradac2004,Limousin2007,Jauzac2015a,Jauzac2016b,Jauzac2017a}.


Another effect of the high mass density of galaxy clusters at large radii is to boost the strong-lensing cross-section of individual galaxies (in particular the ones at or around the cluster redshift), increasing the number of galaxy-galaxy lensing (GGL). Indeed, \citet{Limousin2007} identified three such lenses within 2\arcmin of the core of the massive cluster Abell\,1689, compared to the much lower probability of occurrence of GGL in blank fields (e.g. 10\,deg$^{-2}$, \citealt{Faure2008}). 

The presence of a massive galaxy cluster will locally affect the observed positions of multiple images in a GGL system. Perturbed GGLs are a sign of the effect of the lens environment \citep{Limousin2010}.  \citet{HTu2008} demonstrated how GGL events in cluster fields can be used as direct probes of the radial slope of the cluster density profile (up to $\sim$400\,kpc radius). 
The \emph{Cluster Lensing And Supernovae Survey with Hubble} (CLASH, \citealt{Postman2012}) observed a sample of 25 massive galaxy clusters with the \emph{Hubble Space Telescope} (HST) from the ultra-violet (UV) to the near-infrared (NIR), to study their gravitational lensing properties. This combination of the high-resolution from space with information on colours is perfectly suited to identify GGLs in the cluster outskirts.

In this paper we present a catalogue of candidate GGLs selected in all CLASH fields through visual inspection of the Hubble images. We perform strong-lensing mass reconstructions for three of them, detected in the RXJ2129, MACS\,J0329 and MACS\,J1149 clusters, suitable to probe the cluster mass profiles at large radii, i.e. outside the strong-lensing region, and for which redshift estimates for the lenses and the sources are available. The paper is organised as follows: in Sect.~\ref{sec:observations} we detail the GGL sample selection and the observations at hand; in Sect. \ref{sec:modelling} we present our modelling and results for three GGLs; in Sect. \ref{sec:discussion} we discuss our results and put them in perspective, e.g. GGLs measurements relative to weak-lensing measurements.  

Throughout the paper, we give the magnitudes in the AB system and assume the standard $\Lambda$CDM model with the following cosmology: $\Omega_{m}$=0.3, $\Omega_{\Lambda}$=0.7, and $H_{0}=70\ km\ s^{-1}\ Mpc^{-1}$.

\section{Observations and sample selection}
\label{sec:observations}

We present here the observations and datasets used for our analysis. The  identification of GGLs is based on the inspection of high-resolution HST images from the CLASH program. 

\subsection{Photometric data and GGL selection}

\subsubsection{HST imaging data}

Each cluster was observed with HST in 16 pass-bands, from UV ($\sim$200~nm) to NIR ($\sim$1600~nm) using the \emph{Wide Field Camera 3 }(WFC3) UVIS/IR and the \emph{Advanced Camera for Survey} (ACS). We used the publicly released CLASH images with a pixel scale of 30 mas retrieved from the MAST archive\footnote{\url{https://archive.stsci.edu/missions/hlsp/clash/}}. In the case of MACS\,J0416, MACS\,J0717, MACS\,J1149 and AS\,1063, we used HST images obtained with the \emph{Hubble Frontier Fields} program \citep[HFF;][]{Lotz2017}\footnote{\url{https://archive.stsci.edu/missions/hlsp/frontier/}}, as they supersede the CLASH images in depth near the cluster centre.

\subsubsection{GGL identification}

Several dedicated codes have been developed to perform an automatic detection of gravitational arcs and arclets in wide-field images \citep[e.g. \textsc{arcfinder} and \textsc{yattalens};][]{Seidel2007,Sonnenfeld2017}. Because of the small number of clusters with high-resolution imaging, we preferred to use visual inspection instead. This gives us more flexibility to extend the search in the outskirts of the images where the sky coverage varies from filter to filter. More importantly, we do not focus on a specific lensing configuration (Einstein ring or giant arc) as for the majority of automatic detection codes, and include compact (unresolved) images as well. This visual inspection is not an issue as the completeness of our sample is not necessary for our study.

We focus our search on bright galaxies in the outskirts of the clusters for which strong-lensing models of the cores are available \citep[e.g. ][]{Ebeling2009,Richard2011,Zitrin2012,Zitrin2015}. 
Candidate GGLs were selected in combined-colour images of the clusters using the ($F475W$-$F606W$-$F850LP$) filter combination, or ($F435W$-$F606W$-$F814W$) when HFF images are being used. We also make use of the near-infrared bands ($F606W$-$F105W$-$F160W$) at the cluster cores.
The selection is based on the similarity in colour, morphology and position of the lensed images around bright galaxies. All the GGL candidates are then carefully examined in all HST bands in which they appear, to confirm or discard the strong lensing hypothesis.
A selection of 24 GGL candidates is presented in Table~\ref{tab:GGLcat} and Fig.~\ref{fig:mosaique}.  
Unsurprisingly, our selection detects well-known GGLs. For example the \emph{Dragon Kick} from \citet{DragonKick} or the system \emph{ID14} in \citet{Vanzella2017} in MACS\,J0416.

Considering the importance of GGL events in the outskirts of the clusters, we choose to focus for the rest of the paper on the most interesting GGLs satisfying the following selection criteria:

\textbullet an angular separation from the BCG larger than 80\arcsec ; 

\textbullet plausible lensing configuration from visual inspection (having noticeable multiple images well separated from the lens) ;

\textbullet single, bright galaxy lenses which do not belong to a galaxy group.

\begin{figure*}
\centering
\includegraphics[width=0.33\linewidth]{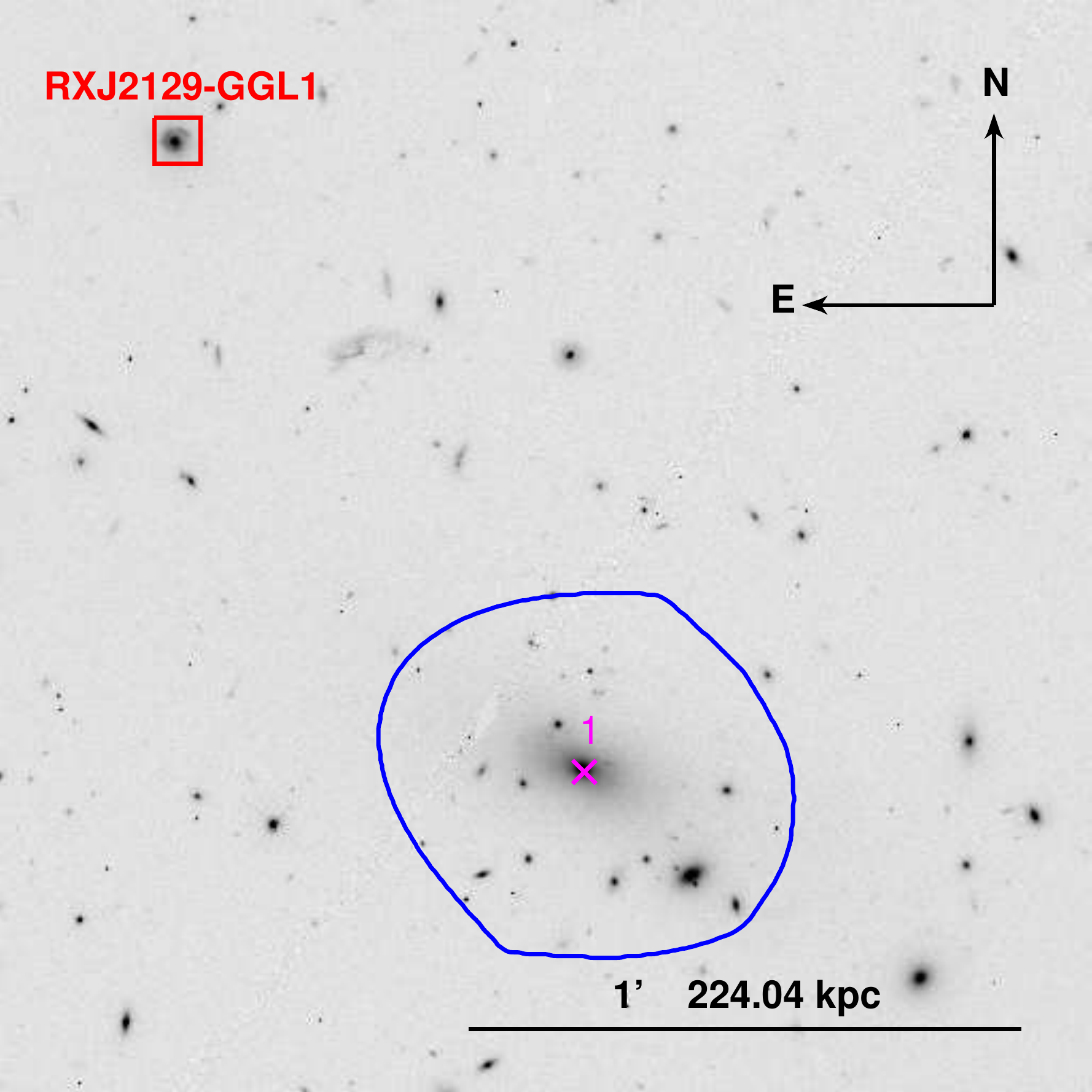}
\includegraphics[width=0.33\linewidth]{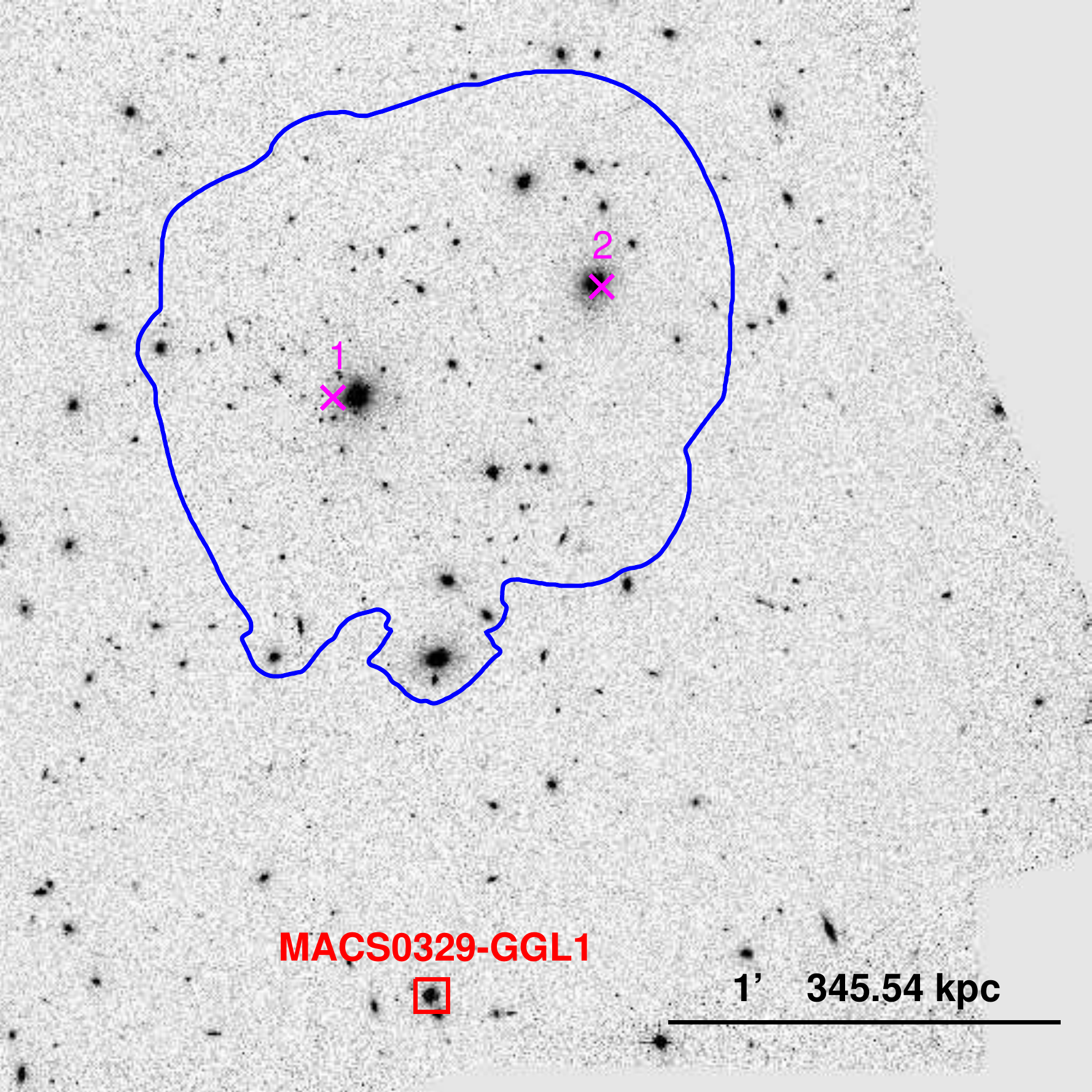}
\includegraphics[width=0.33\linewidth]{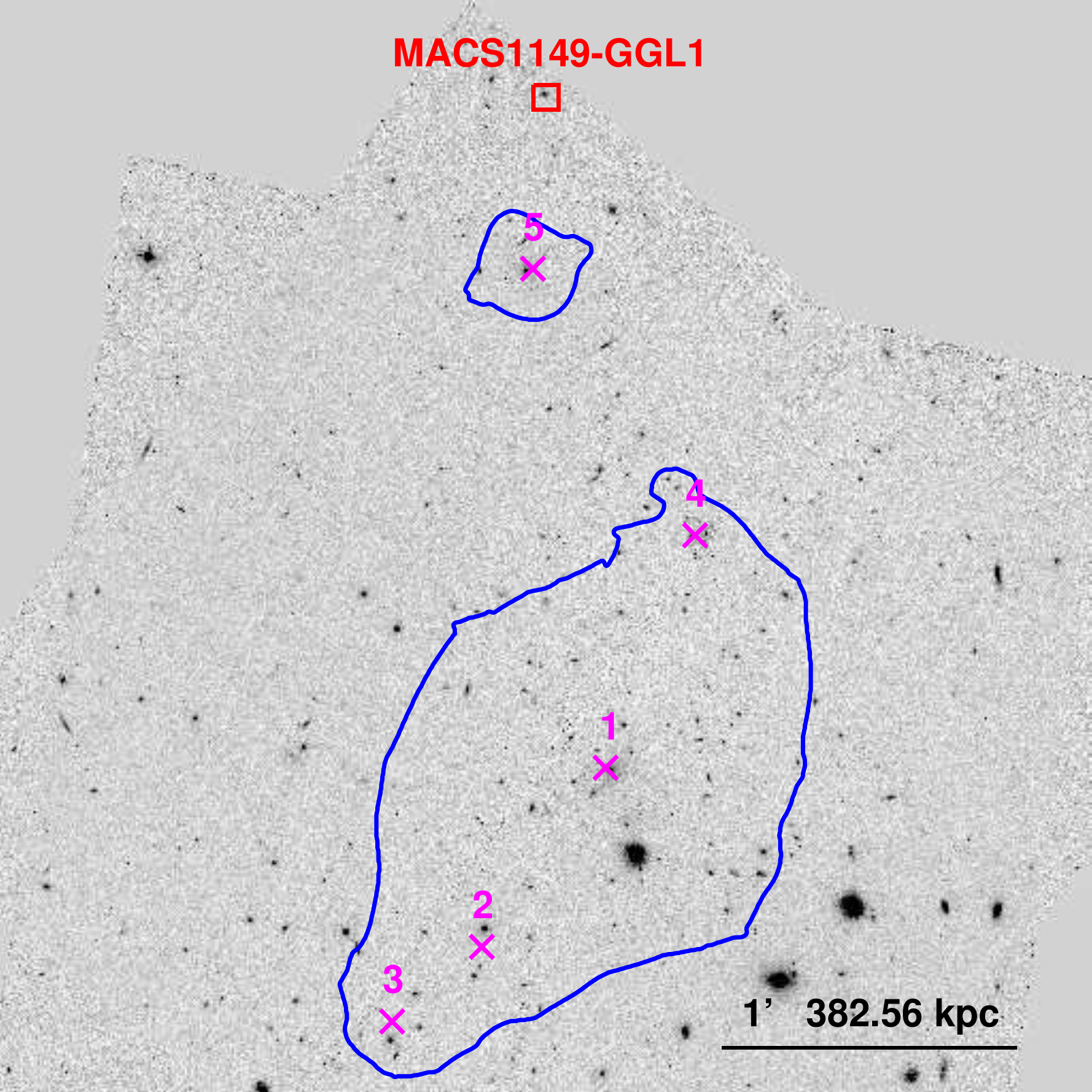} \\
\includegraphics[width=0.328\linewidth]{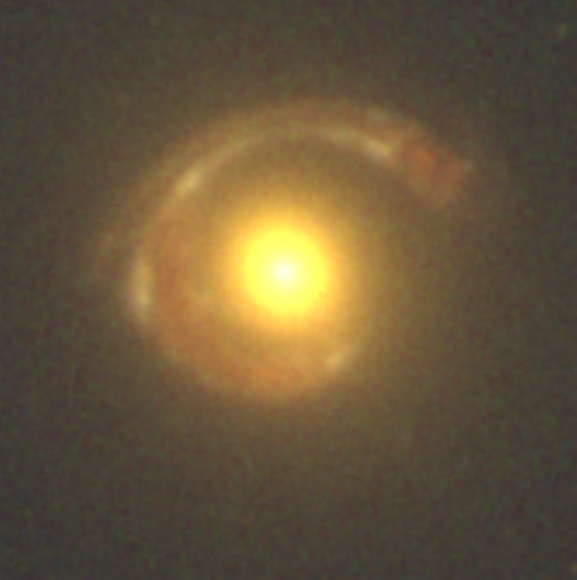} 
\includegraphics[width=0.33\linewidth]{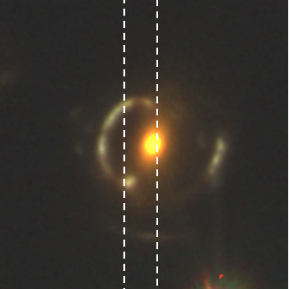} 
\includegraphics[width=0.33\linewidth]{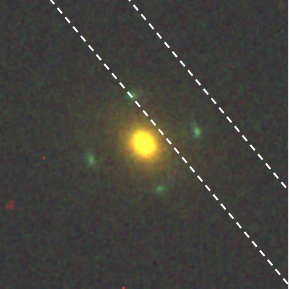} 
\caption{HST images of the three clusters hosting the three GGL candidates that are at the core of this paper. \emph{Top:} a view of the clusters in the $F775W$-band. The location of the GGL in each cluster is highlighted with a red box. The blue contours delineate the multiple image regions expected for sources at $z=6$. \emph{Bottom:} From left to right: RXJ2129-GGL1 (a.k.a the \emph{Snail}), MACS\,J0329-GGL1 and MACS\,J1149-GGL1. The images are $5\times 5$\,arcsec$^{2}$. For the first two images, the blue channel combines $F435W$ and $F475W$ filters, the green one $F606W$ and $F625W$, and the red one $F775W$-, $F814W$- and $F850LP$-bands. For the third stamp, $F435W$, $F606W$ and $F775W$ are being used for the blue, green and red channels respectively. North up and East is left. The dashed white lines on the bottom middle and right panels indicate the slits positioning for the spectroscopy.}
\label{fig:selection}
\end{figure*}

This selection provides us with three GGL candidates highlighted in the bottom panel of Fig.~\ref{fig:selection}, with their characteristics listed in Table~\ref{tab:Selected_GGL}: 

(i) RXJ2129-GGL1 is being quadruply-lensed by an elliptical galaxy. Its four extended multiple images are seen to spiral around the lens. This elliptical and spiral-like configuration led us to refer to it as the \emph{Snail}.

(ii) MACS\,J0329-GGL1 is a system surrounding a central elliptical galaxy. It consists of an extended arc to the East and a smaller arc to the West. We note that the distances of the two arcs from the lens are unusually different. The colours of the arcs components being the same suggest that there is a single background source. 

(iii) MACS\,J1149-GGL1 is being lensed by an elliptical galaxy, and forms an almost perfect Einstein cross: the four images are nearly symmetric with a small angle from a perfectly perpendicular cross. 

\begin{table*}
\caption{The three GGLs selected in this study. From left to right: ID, coordinates (J2000) of the centre of the lens, redshift of the cluster, redshift of the galaxy lens and of the source, distance to the BCG, the magnitude measured in the $F775W$-band of the lens and the source.}
\begin{tabular}{l c c r r r r r r}
\hline
ID & $\alpha$ & $\delta$ & z$_{c}$  & z$_{l}$ & z$_{s}$ & $d_{BCG}$ &  F775W & source F775W\\
& & & & & & [arcsec]  & [mag] & [mag]$^{*}$\\
\hline
RXJ2129-GGL1 & 322.4287798 & 0.1080707 & 0.235 & 0.255$^{+0.033}_{-0.021}$ $^{a}$ & 1.61$^{+0.37}_{-0.31}$ $^{a}$ & 81.0 & 17.58$\pm$0.01 & 21.02$\pm$0.04 \\
MACS\,J0329-GGL1 & 52.4201304 & -2.2216321 & 0.45 & 0.3835$^{b}$ & 1.112$^{b}$ & 92.0 & 19.59$\pm$0.01 & 19.88$\pm$0.01 \\
MACS\,J1149-GGL1 &  177.40.28.221 & 22.43.66292 & 0.544 & 0.542$^{b}$ & 1.806$^{b}$ & 137.9 & 20.22$\pm$0.06 & 22.52$\pm$0.02\\
\hline
\multicolumn{9}{l}{$^{a}$Photometric redshift with 2$\sigma$ error (Sect.~\ref{sec:photoz}); $^{b}$ Spectroscopic redshift (Sect.~\ref{sec:specz}); $^{*}$ observed magnitude}
\end{tabular}
\label{tab:Selected_GGL}
\end{table*}

\subsubsection{Deblending}
\label{sec:deblending}

The lens and multiple images in MACS\,J0329-GGL1 and MACS\,J1149-GGL1 are well-separated. However, in the case of RXJ2129-GGL1 the lens is contaminating the source. In order to obtained a precise photometry, we modelled the lens and subtracted it from the image, using the {\sc galfit} \citep{2011ascl.soft04010P} software. 

\begin{figure}
\centering
\includegraphics[width=.15\textwidth]{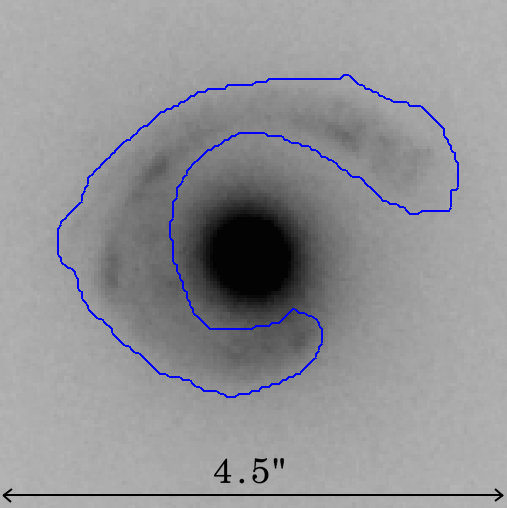}\hfill\includegraphics[width=.15\textwidth]{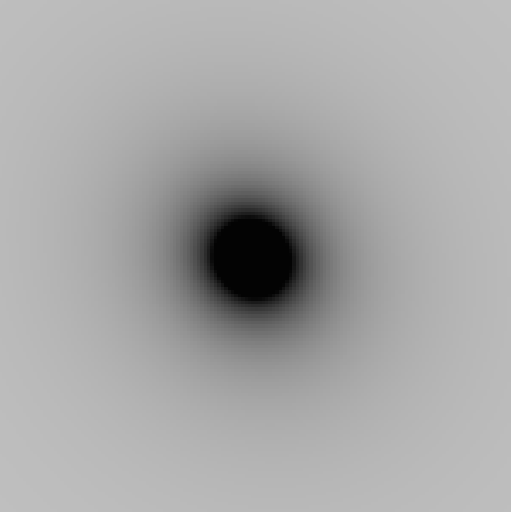}\hfill\includegraphics[width=.15\textwidth]{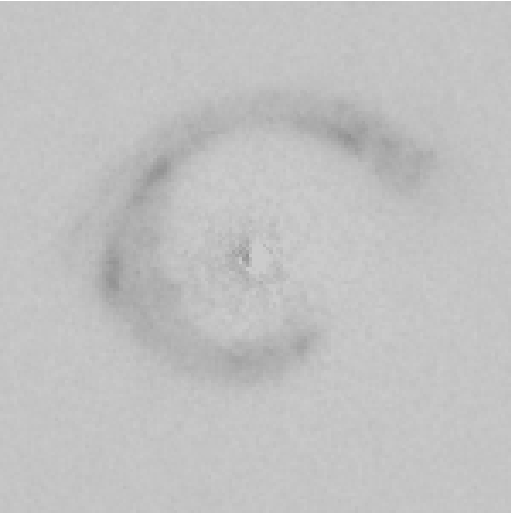}\hfill
\caption{Deblending method on RXJ2129-GGL1. \emph{Left pannel:} The initial image as seen in the $F814W$-band. The blue contour highlights the shape used for the mask (see \ref{sec:deblending}), and also used for the aperture photometry (see section~\ref{sec:phot})\emph{Middle panel:} The sersic model of the lens fitted by {\sc galfit}. \emph{Right panel:} The residual image.}
\label{fig:deblending}
\end{figure}

The input files are first generated by {\sc galapagos} \citep{2012MNRAS.422..449B}. We then manually define the input mask to reject all pixels belonging to the source (blue contour shown in the left panel of Fig.~\ref{fig:deblending}) in the modelling of the lens. {\sc galfit} fits the lens with a Sersic profile using the 8 different available pass-bands between $F390W$ and $F850LP$ (Table \ref{tab:photometry}) where the lens is fully detected. During the modelling, parameters such as position ($x$,$y$) and shape (radius, axis ratio and position angle) are assumed to be constant with wavelength. The Sersic index can linearly evolve with wavelength, and the magnitude is considered as a free parameter. The residual image is shown in the right panel of Fig.~\ref{fig:deblending}.

\subsubsection{Photometry}
\label{sec:phot}

Photometric catalogues are publicly available for all CLASH clusters as part of the delivered high level science products\footnote{\url{https://archive.stsci.edu/missions/hlsp/clash/}}, providing positions, shapes, magnitudes and photometric redshifts of the extracted objects. These catalogues are used to derive the photometry in MACS\,J0329-GGL1.

None of the lensed images in MACS\,J1149-GGL1 multiple images is detected in this catalogue. We used \textsc{sextractor} \citep{Bertin1996} in order to get a clean photometry for these images in all the bands where the GGL appears ($F435W$, $F606W$, $F625W$, and $F775W$). The combined $F775W$-band magnitude is provided in Table~\ref{tab:Selected_GGL}.

To measure the source magnitudes for RXJ2129-GGL1 we use the residual image presented in Sect.~\ref{sec:deblending} (Right panel of Fig.~\ref{fig:deblending}). Due to the complex morphology of this source, we use the manually defined aperture (blue contours in Fig.~\ref{fig:deblending}) to measure the source flux, and then remove the background previously estimated in an outer annulus ($2.1-2.4\arcsec$). 
In the case of the lens we use {\sc galfit} to fit a Sersic profile to the lens and get the magnitudes, a mask to hide the source flux has been applied. Magnitudes measured by {\sc galfit} are listed in Table~\ref{tab:Selected_GGL} and Table~\ref{tab:photometry} .

For the modelling part detailed in Sect.~\ref{sec:modelling}, we need the geometrical parameters (centroid, $\alpha_{c}$ and $\delta_{c}$, ellipticity, $e_{c}$, position angle, $\theta_{c}$) and the luminosity of the cluster members. For MACS\,J1149 and RXJ2129, we use the galaxy catalogues from \citep{Jauzac2016b} and \cite{Richard2010} respectively. We incorporate the photometry of the new CLASH images in the RXJ2129 catalogue, using the $F160W$-band. For the galaxies not appearing in the WFC3 field of view, we use ACS/$F814W$ and apply a mean ($F160W$-$F814W$) colour estimated with the  \citet{Coleman} empirical template for elliptical galaxies. We also use the geometrical parameters ($\alpha_{c}, \delta_{c}$, $e_{c}$, $\theta_{c}$) measured in the $F814W$ band for the RXJ2129 cluster members catalogue.

In the case of MACS\,J0329, we select the cluster members following the red sequence technique on a ($F606W$-$F814W$) \emph{vs.} $F814W$ colour-magnitude diagram. 
We chose a limiting magnitude $F814W=23$ and a colour width of 0.3\,magnitude for the red sequence (above three times the photometric uncertainties). We incorporate the $F160W$ photometry when galaxies are visible in this pass-bands.
Finally we add the geometrical parameters ($\alpha_{c}, \delta_{c}$, $e_{c}$, $\theta_{c}$) measured in the $F814W$-band to the catalogue.

\subsection{Redshift estimates}

\subsubsection{Spectroscopic redshift}
\label{sec:specz}


All CLASH clusters have been extensively covered with the \emph{VIsible MultiObject Spectrograph} (VIMOS,  \citealt{VIMOS2003}) on the \emph{Very Large Telescope} (VLT), as part of the ESO program 186.A-0798 (PI: Rosati, \citealt{Rosati}). We looked at all the masks covering the three studied clusters, and found that MACS\,J0329-GGL1 had been targeted for one 1125\,sec exposure obtained with the \textsc{MR} medium resolution ($R=580$) grism during the night of Dec. 01 2012. The slit position is presented in the bottom middle panel of Fig~\ref{fig:selection}.

\begin{figure*}
\centering
\includegraphics[width=\linewidth]{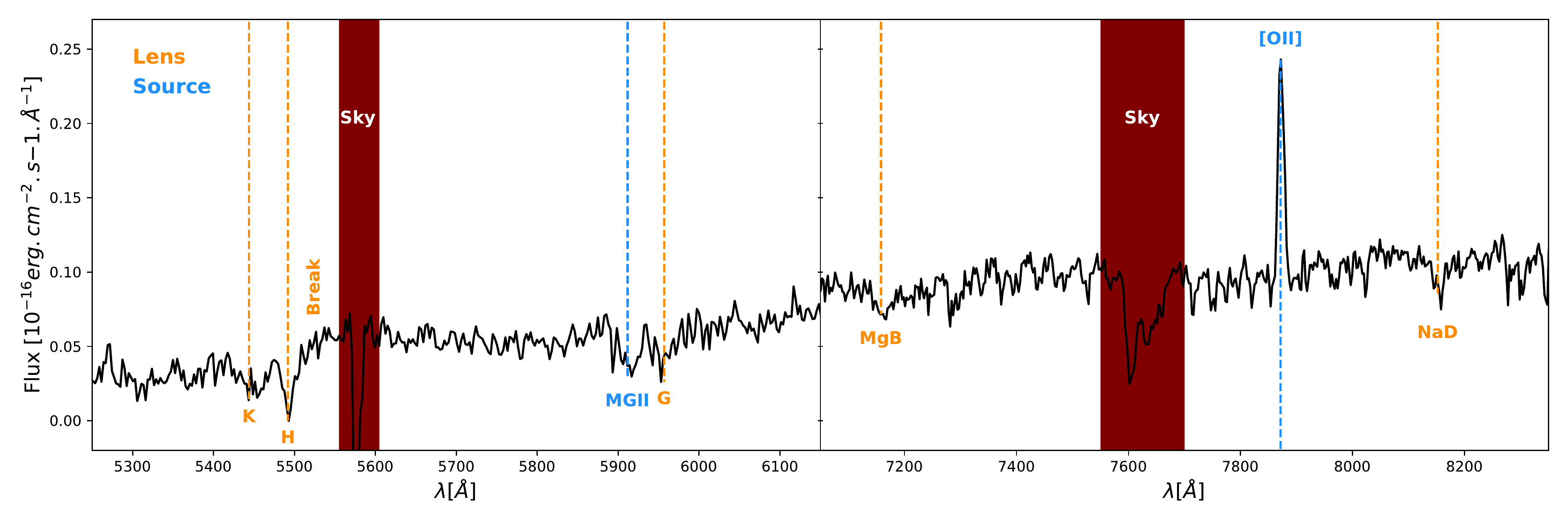}
\caption{VIMOS extracted spectrum of both MACS\,J0329-GGL1 lens and source. In orange, the lines of the lens with a redshift z$_{l}$~=~0.3835, and in blue, the line of the source with a redshift z$_{s}$~=~1.112. We identify the K, H, G MgB and NaD absorption lines for the lens and its Balmer break. For the source, we observe the [\ion{O}{ii}] emission lines and the \ion{Mg}{ii} absorption lines.}
\label{fig:spectrum}
\end{figure*}

The spectra were extracted using the VIMOS pipeline v2.9.16. Following the instruction in the manual v6.8\footnote{\url{ftp://ftp.eso.org/pub/dfs/pipelines/vimos/vimos-pipeline-manual-6.8.pdf}}, we performed standard reduction with the new recipes for bias removal, flat-field correction, wavelength calibration, sky subtraction and used observations of spectroscopic standard stars to derive the flux calibration.

\begin{figure}
\centering
\includegraphics[width=\linewidth]{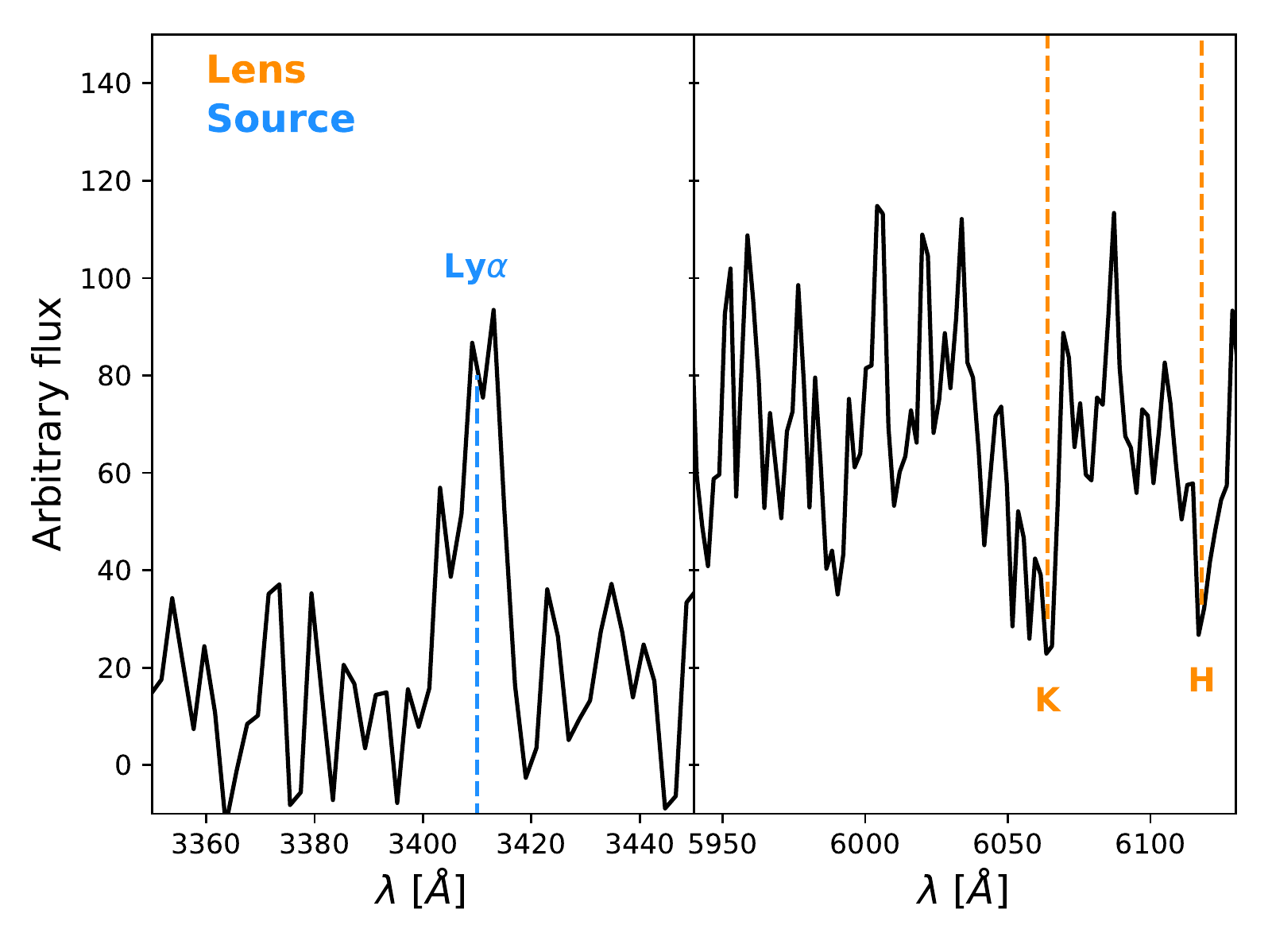}
\caption{Spectrum of the MACS\,J1149 GGL. A Ly$\alpha$ emission is detected at $\lambda=3410$\AA from the background source. Ca absorption line from the lens are present at $\lambda_{K}=6066$ \AA\ and $\lambda_{H}=6120$\AA, at the edge of the LRIS blue arm spectral coverage. This correspond to z$_{s}=1.806$ and $z_{l}=0.542$ respectively.}
\label{fig:specm1149}
\end{figure}

The extracted spectrum of the galaxy in the medium resolution grism is presented in Figure~\ref{fig:spectrum}. We identify the presence of K, H, G and NaD absorption lines and a Balmer break at a redshift $z_{l}=0.3835$. We also note an emission line that does not match the lens redshift. We identify it as an [\ion{O}{ii}] emission line belonging to the source at redshift $z_{s}=1.112$. This redshift is consistent with additional absorption lines of \ion{Mg}{ii} in the continuum.


%

 A spectrum of the western image of MACS\,J1149-GGL1 was obtained with the LRIS instrument (\citealt{Oke1995}, \citealt{Steidel2004}) on the Keck I telescope. The position angle was 40$^{\circ}$ and the slit width was 1.0$''$ (see Fig~\ref{fig:selection}) and the airmass ranged from 1.03-1.12.  Three exposures of 27 minutes each were taken for a total exposure time of 81 minutes.  
%

The extracted spectrum is presented in Fig.~\ref{fig:specm1149}. Spectral features are detected from both the lens and background source. Strong Ly$_{\alpha}$ is found in emission at $\lambda$=3410\AA\ corresponding to a redshift $z_{s}=1.806$. In the red part of the spectrum we observe K and H absorption lines of the lens galaxy (not centred in the slit) at wavelengths of 6066\AA\ and 6120\AA\ respectively. This corresponds to a redshift z$_{l}=0.542$ for the lens, in agreement with the cluster redshift.

\subsubsection{Photometric redshifts}
\label{sec:photoz}

\begin{figure}
\centering
\includegraphics[width=\linewidth]{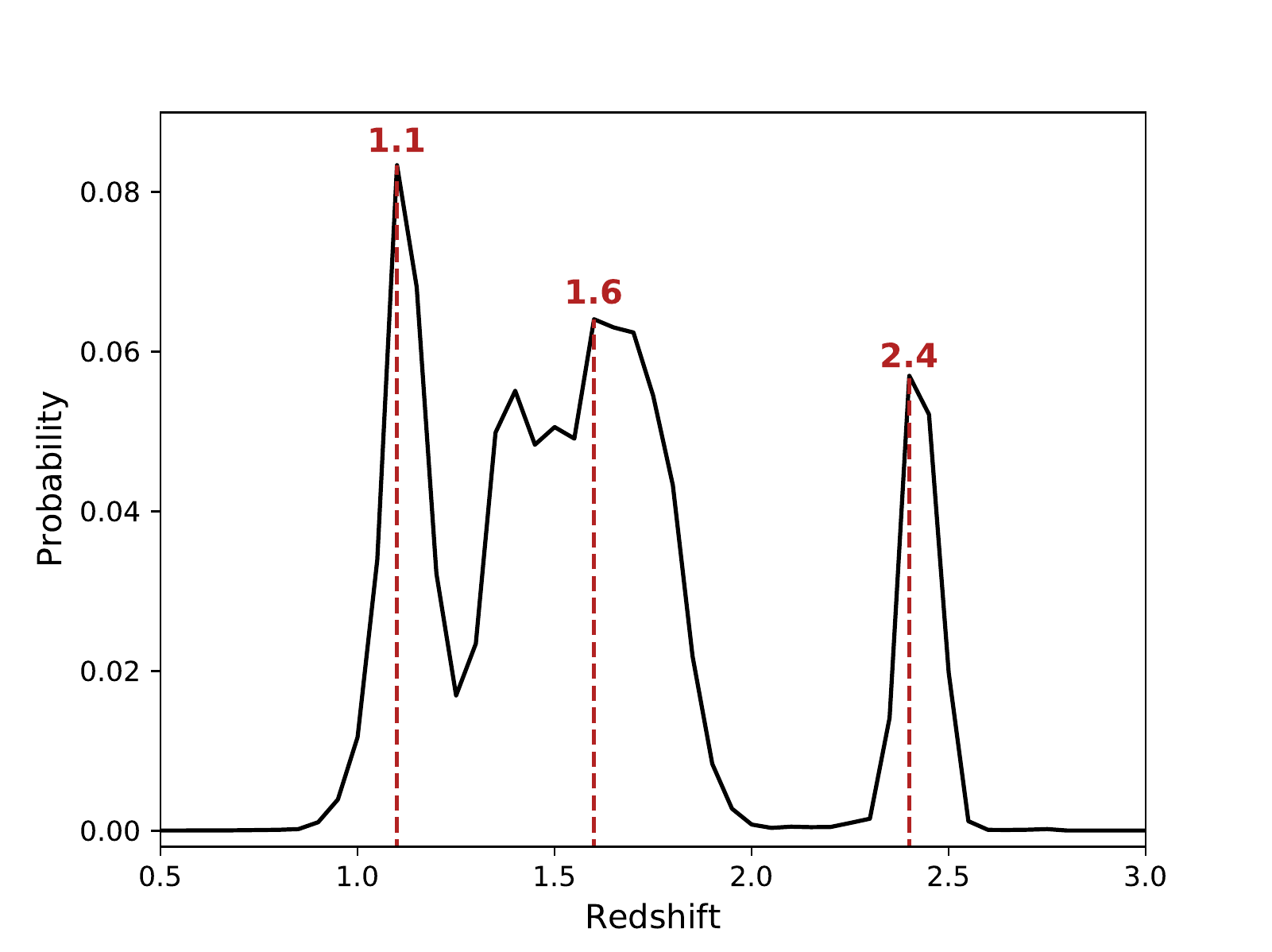}
\label{fig:pdf}
\caption{Probability density function of the \emph{Snail} source photometric redshift. We note three maxima, located at redshift $z=1.1$, $z=1.6$ $z=2.4$. }
\end{figure}

For the RXJ2129-GGL1 system, we used {\sc hyperz} \citep{2011ascl.soft08010B} to fit the spectral energy distribution (SED) and estimate photometric redshifts for the lens and the source. To fit the SED we used models made from \cite{2003MNRAS.344.1000B} with an initial mass function (IMF) from \cite{1955ApJ...121..161S} and a metallicity of $0.02\,Z_{\odot}$, and with the reddening law of \cite{2000ApJ...533..682C} we allowed AV to be in the range $[0.0-3.0]$. {\sc hyperz} provides the probability distribution of the photometric redshift of the system. It shows three maxima at $z=1.1$, $z=1.6$ and $z=2.4$ (Fig~\ref{fig:pdf}). 

Based on the physical parameters derived on the lens during our modelling (Sect.~\ref{sec:modelling}) the redshift solution $z=1.6$ is preferred, and is given with its associated error in Table~\ref{tab:Selected_GGL}. We discuss this assumption later in Sect.~\ref{sec:photozdiscuss}.

\section{Models}
\label{sec:modelling}

We build parametric models of the mass distribution of the GGLs in order to reproduce the observed lensing configurations. Models have varying complexity in order to test different assumptions of the impact of the environment. The same methodology was applied to each GGL for constructing the models and analysing the results.

\subsection{Methodology}

We use the software \texttt{Lenstool} \citep{Kneib1993,Jullo2007} to optimise parametric models of the mass distribution in each system. 
\texttt{Lenstool} uses the observed positions of multiple images as constraints. For a set of mass parameters and a given system of multiple images, it computes the barycentre of all positions in the $source$ plane. It then lenses this location back into the $image$ plane. The model parameters are sampled using a Monte-Carlo Markov-Chain (MCMC) and optimised through a $\chi^{2}$ minimisation using the distances between the observed and model-predicted positions of the multiple images.

The three GGLs are dominated by a massive central galaxy lens. The mass distribution of the lens galaxy is usually well-described with a single parametric potential but the effect of its (generally unknown) environment is included in the form of a constant external shear field (\citealt{Schechter97},\citealt{Dye2007},\citealt{Wagner2016},\citealt{Wong2017}). Here we know that the environment of each GGL is certainly dominated by the nearby massive galaxy cluster.

The mass distribution is modelled by a superposition of mass components describing galaxy- and/or cluster-scales. These gravitational potentials are described by double Pseudo-Isothermal Elliptical profiles (dPIE, \citealt{PIEMD2007}). This distribution is described by the following parameters: 

\textbullet \ the geometrical parameters (central position $\alpha_{c}, \delta_{c}$, ellipticity and position angle $e_{c}$, $\theta_{c}$),

\textbullet \ the central velocity dispersion, $\sigma_{0}$,

\textbullet \ a cut radius, $r_{cut}$,

\textbullet \ a core radius, $r_{core}$.

Four models are constructed for a given GGL, each model getting a higher level of complexity than the previous depending on the assumption used on the environment. We start by only modelling the single central galaxy lens, and finally the whole cluster and GGL are constrained together. 

We adjust for each model the parameters to optimise and the range of values. The results presented in this work are the best models, with the lowest $\chi^{2}$, with the parameters presented in Appendix~\ref{appendixB}.

\subsubsection*{Model \texttt{I}: single galaxy lens}

In this model we only consider the lens of the GGL as the deflector, ignoring the effect from other lenses. Only the $\sigma_{0}$ of the mass component is optimised, its ellipticity and position angle are set to the ones of the light measured in Sect.~\ref{sec:phot}. Due to its degeneracy with $\sigma_0$ \citep{Richard2010}, r$_{cut}$ is fixed to a typical value of 50\,kpc. This hypothesis is further discussed in Sec.~\ref{sec:degen}. r$_{core}$ is fixed to 0 as it does not have an impact on the lensing effect.

\subsubsection*{Model \texttt{II}: single galaxy lens and external shear}

In this model, we use the same parametrisation as model I for the galaxy, and assume the lensing contribution of the environment surrounding  the GGL is well modelled by adding a constant external shear. The magnitude $\gamma$ of this shear and its orientation $\theta$ are free additional parameters like $\sigma_{0}$ of the lens. The shear magnitude is given for a D$_{LS}$/D$_{OS}$, ratio of angular distances between the lens and the source and between the observer and the source respectively, equal to 1.

\subsubsection*{Model \texttt{III}: cluster and GGL}
\label{sec:clustermodel}

This model includes a full optimisation of the cluster and the GGL system. Cluster size potentials are being optimised with a fixed r$_{cut}$=1000\,kpc, but their position, $\sigma_{0}$, ellipticity, position angle and r$_{core}$, are free to vary. The BCG as well as the lens of the GGL system are being modelled by a dPIE potential, setting $\sigma_{0}$ as a free parameter. 

With a sufficient number of constraints, the r$_{cut}$ of the BCG can be optimised. Cluster members are being modelled by individual galaxy-size potentials, but to limit the number of parameters we assume they follow the Faber-Jackson scaling-relation \citep{FJ76} as described in \citet{Richard2010}:

\begin{equation}
\sigma_0=\sigma^{*}_{0}\,\Big(\frac{L_{\rm F160W}}{L^{*}_{\rm F160W}}\Big)^{1/4}
\end{equation}

\begin{equation}
r_{cut}=r_{cut}^{*}\,\Big(\frac{L_{\rm F160W}}{L^{*}_{\rm F160W}}\Big)^{1/2}
\end{equation}

\begin{equation}
r_{core}=r_{core}^{*}\,\Big(\frac{L_{\rm F160W}}{L^{*}_{\rm F160W}}\Big)^{1/2}
\end{equation}

This relation links the $F160W$-band luminosity, L$_{F160W}$, to a L$^{*}_{F160W}$, and scales the mass parameters of the cluster members to the ones of the standard galaxy ($\sigma^{*}_{0}$, r$^{*}_{cut}$, r$^{*}_{core}$).
The luminosity of the standard galaxy is computed following the results of \citet{Lin2006} as in the work of \citet{Richard2010}.
We optimise the $\sigma^{*}_{0}$ and fix $r^{*}_{cut}$ at 45\,kpc and $r^{*}_{core}$ at 0.15\,kpc.

All the multiple image systems are included as constraints to this model. In the case of an unknown redshift of the source, the redshift is included as a free parameter of the model.

\subsubsection*{Model \texttt{IV}: cluster only}

This model is similar to the previous one, but the GGL multiple images are not used to constrain the model. The lens of the GGL, when in the cluster, is included and assumed to follow the scaling relation described before. We use this model as a point of comparison with the model \texttt{III} to estimate the impact of the GGL constraints on the cluster mass distribution.

\subsubsection*{Analysis of the results}

For the models \texttt{I}, \texttt{II} and \texttt{III}, we use as a comparison parameter the root mean square (RMS) of the distance between the observed and predicted position of the multiple images. The RMS for all three clusters and models are listed in Table~\ref{tab:rms}, and further discussed in Section~\ref{sec:discussion}.

We also compare the produced shear by the models \texttt{II}, \texttt{III} and \texttt{IV}. The result of the shear optimisation is scaled with the D$_{LS}$/D$_{OS}$ factor of the GGL for the model \texttt{II}. For the two others, the shear is measured by making a shear map at the position of the GGL after subtracting it from the models. We construct 5'$\times$5' maps of 50 pixels across for the two components of the shear, $\gamma_{1}$ and $\gamma_{2}$ \citep[$\gamma \equiv \gamma_{1} + i\gamma_{2}$ ; see][]{Bartelmann2001}. We then measure their mean values.

From these values of $\gamma_{1}$ and $\gamma_{2}$, we then compute the magnitude and the orientation of the shear. 
We apply the same methodology to all the realisations of each model. We can then measure the scatter in both shear magnitude and orientation. To compute contours containing 68.3\% and 95.4\% of all the points, we used a Gaussian kernel density estimation with a bandwidth selected using \citet{Scott1992}'s rule of thumb. 
The results is a contour map of the shear versus its orientation is produced (see Fig~\ref{fig:shearrxj}, \ref{fig:macs0329_shear} and \ref{fig:m1149shear}). 

\subsection{RXJ2129}

\label{paramrxj}
Our model of RXJ2129 (z$_{c}$~=~0.235) is based on the one presented by \cite{Richard2010}. This model includes 39 cluster galaxies, comprising both the BCG and the central galaxy lens in RXJ2129-GGL1. \cite{Richard2010} used a triply-imaged system near the BCG with a known spectroscopic redshift, $z=1.965$. Since then, this redshift has been revised to $z=1.522$ \citep{Belli2013}. We include in our model two multiply-imaged systems from \cite{Zitrin2015}: systems \#3 and \#5. For system \#5, we only use images 1 and 2. For both systems the redshift is included as a free parameter.

\begin{table*}
\centering
\begin{tabular}{l c c c c c c c c r}
\hline
ID & F390W & F435W & F475W& F606W & F625W & F775W & F814W & F850LP & Photo-$z$ \\
\hline
 \multicolumn{2}{l}{RXJ2129} \\
Lens & 20.55$\pm$0.01 & 20.10$\pm$0.01 & 19.41$\pm$0.01 & 18.26$\pm$0.01 & 18.05$\pm$0.01 & 17.58$\pm$0.01 & 17.50$\pm$0.01 & 17.24$\pm$0.01 & 0.245$^{+0.086}_{-0.019}$\\
Source$^{a}$ & 22.86$\pm$0.12 & 22.48$\pm$0.07 & 22.27$\pm$0.07 & 21.78$\pm$0.02 & 21.61$\pm$0.03 & 21.02$\pm$0.04 & 20.79$\pm$0.03 & 20.32$\pm$0.02 &  1.61$^{+}_{-}$\\
\hline
\multicolumn{8}{l}{$^{a}$ Photometry combines all multiple images together.}
\end{tabular}
\caption{Photometry for the GGL of RXJ2129 in all the available bands and the computed photo-$z$.}
\label{tab:photometry}
\end{table*}

The \emph{Snail} is a GGL located North-East of RXJ2129 core, at a distance of 81\arcsec from the BCG (see Figure~\ref{fig:selection} ; Table~\ref{tab:Selected_GGL}). As the image in Fig.~\ref{fig:selection} shows, one can see four multiple images around the central elliptical galaxy. Their positions are listed in Table~\ref{tab:rxjmul}. We note that all images are close to the lens, leading to a contaminated photometry. That problem can be solved by subtracting the central galaxy in all the bands and is discussed in Section~\ref{sec:deblending}.

The photometry of the images after subtraction is given in  Table~\ref{tab:photometry}. The photometry of the source is measured trough an aperture that is covering all the multiple images (see Fig.~\ref{fig:deblending}). We use that photometric catalogue to compute a photometric redshift for both the lens and the source. In \cite{Richard2010}, the \emph{snail} was considered as a cluster member, the photometric redshift was measured at $z=0.255^{+0.033}_{-0.021}$, consistent within the error bars to the one of the cluster z$_{c}$~=~0.235. 

We can also note from the illustration of Figure~\ref{fig:selection} that the ring of multiple images is being sheared. This shear seems to be perpendicular to the direction of the BCG (see Fig.~\ref{fig:selection}~\ref{fig:rxjpos}).
The best-fit parameters for all RXJ2129 models are given in Table~\ref{tab:rxjparams}.

\subsubsection{Model \texttt{I}}

The lens of RXJ2129-GGL1 was already included in the cluster scaling relations by \citet{Richard2010}. When modelling the GGL with a single galaxy potential, we fix its r$_{core}$ to 0\,kpc and r$_{cut}$ to 64\,kpc. 

The best-fit model predicts the images as they are presented in Fig.~\ref{fig:rxjpos} (yellow diamonds) with an RMS of 0.66\arcsec\ (Table~\ref{tab:rms}). The image at the North of the \emph{snail} is not computed, and two images are predicted on the East. Also the images are all predicted at a similar distance to the lens, indicating a ring-like configuration instead of the observed elliptical configuration. We thus conclude that a single galaxy lens is not sufficient to recover the observed configuration.

\subsubsection{Model \texttt{II}}

\begin{figure}
\includegraphics[width=\linewidth]{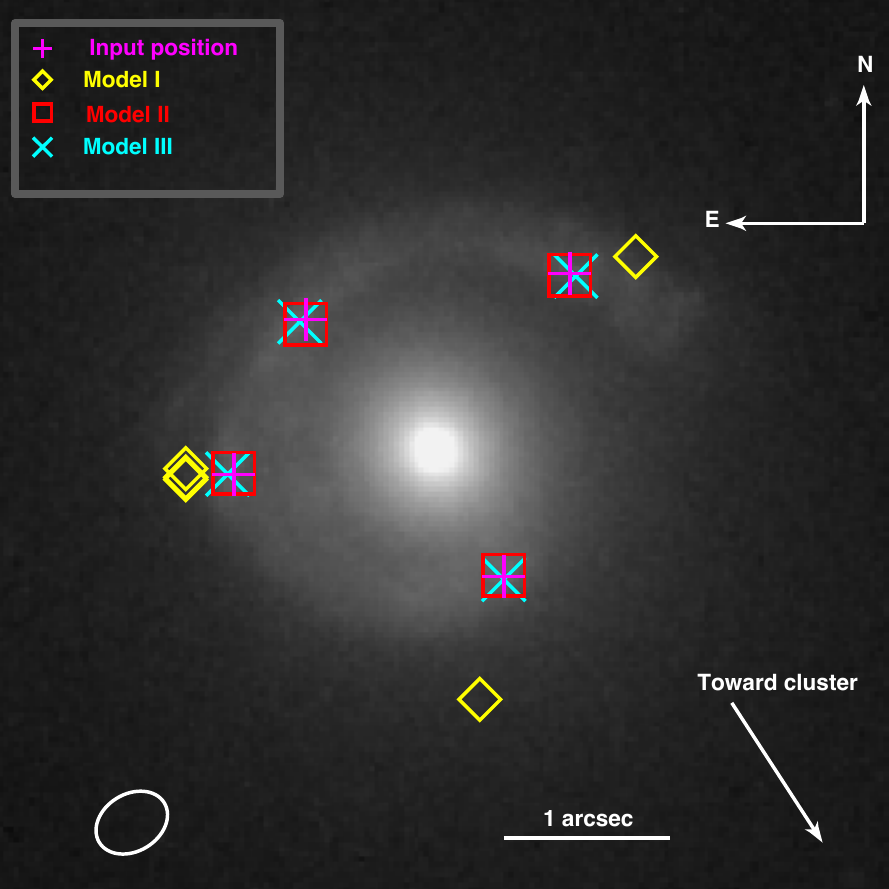} 
\caption{Predicted positions of the multiple image for all models for RXJ2129-GGL1. Magenta crosses are the observed positions, yellow diamonds are the predictions from Model \emph{I}, red boxes are the predictions from Model \emph{II} with the external shear, and the cyan crosses are the predictions from Model \emph{III} which includes the cluster. The white ellipse (lower left corner) is the representation of the external shear on a circle of radius 0.2\arcsec.}
\label{fig:rxjpos}
\end{figure}

The predicted positions of the multiple images when considering an external shear are shown in Fig.~\ref{fig:rxjpos} (red squares). They are in much better agreement with the observed ones as shown by an RMS of 0.02\arcsec\ (see Table~\ref{tab:rms}). The best model gives an external shear of amplitude $\gamma$~=~0.15$^{+0.04}_{-0.03}$, and angle from the West direction of $\theta$~=~31.5$^{+3.3}_{-3.2}$. That orientation is consistent with the direction toward the centre of the cluster with the BCG being oriented perpendicular to the predicted shear (see Table~\ref{tab:SIS}). 

\begin{figure}
\includegraphics[width=\linewidth]{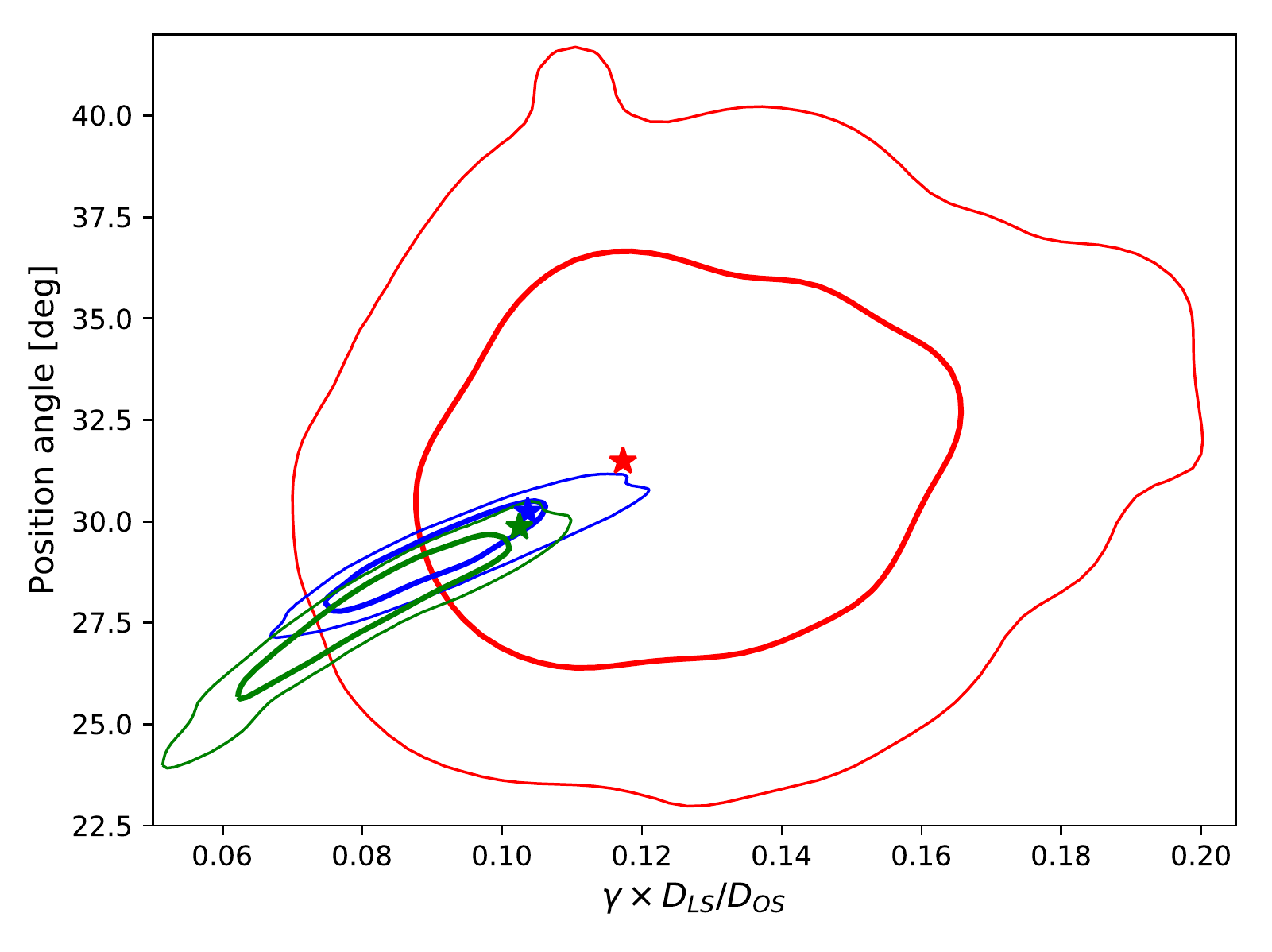} 
\caption{Predicted shear at the position of the \emph{snail} \emph{versus} its orientation: in red the external shear predictions (Model \emph{II}), in green the predictions of the cluster-only model (Model \emph{IV}), and in blue the predictions from Model \texttt{III}. The bold contour represents the 1$\sigma$ limit and the thin one the 2$\sigma$. The stars show the best predictions for each model. The D$_{LS}$/D$_{OS}$ factor is the one of the GGL.}
\label{fig:shearrxj}
\end{figure}
%
%
\subsubsection{Model \texttt{III}}

Here, we model both the \emph{snail} and the cluster. As in \cite{Richard2010}, each cluster galaxies, excepted for the \emph{snail} lens and the BCG, are modelled by a dPIE potential and following the \cite{FJ76} scaling relation. The K-band luminosity was used to scale the parameters of the cluster members. We thus convert the $L^{*}_{K}$ to $L^{*}_{F160W}$ to scale the parameters with the $F160W$-band CLASH magnitudes leading to $m_{F160W}^{*}=17.49$. Finally, only the $\sigma_{0}^{*}$ of the reference galaxy is optimised. The BCG and the \emph{snail} lens are being optimised individually, and only $\sigma_{0}$ is set as a free parameter.

To model the influence of the cluster at large radii, we create a PIEMD halo with a $r_{cut}=1000$\,kpc. \texttt{lenstool} optimises the position of this halo in a box of 5\arcsec centred around the BCG, its orientation, ellipticity, $r_{core}$ and $\sigma_{0}$.

We use as constraints all the multiple images presented in Table~\ref{tab:rxjmul}. For systems \#3 and \#5, we optimise the redshift and only use the positions of the multiple images as constraints. The best fit model gives $z_{s3}$~=~1.49$^{+0.17}_{-0.09}$ and $z_{s5}$~=~0.78$^{+0.05}_{-0.03}$. These results are within the 95\% confidence interval presented by \cite{Zitrin2015}. 

The predicted positions of the \emph{snail} are similar to the one obtained with Model \emph{II} (with an external shear). The resulting RMS is 0.03\arcsec\ \emph{versus} 0.02\arcsec\ for Model \emph{II}. Figure~\ref{fig:rxjpos} shows the predicted positions of the multiple images as cyan crosses.
In Fig.~\ref{fig:shearrxj}, one can see the shear produced by the cluster in this model. The predicted shear from the cluster itself is close in orientation and intensity to the external shear obtained with Model \emph{II}. 

\subsubsection{Model \texttt{IV}}

The lens of the RXJ2129-GGL1 is treated in this model as a cluster member and optimized through the scaling relation. The GGL lens being at the edge of the HST/WFC3 field of view, its photometry in the $F160W$-band is computed as explained in Sec.~\ref{sec:phot}.

The best fit model predicts a redshift of $z_{s3}$~=~1.55$^{+0.17}_{-0.11}$ and $z_{s5}$~=~0.79$^{+0.05}_{-0.04}$ for system \#3 and \#5 respectively. These results are close to the ones from Model \texttt{III}. 
Figure~\ref{fig:shearrxj} shows the shear prediction at the position of the \emph{snail}. We note that the contours predicted by Model \texttt{III} and Model \texttt{IV} are similar, but the one including the GGL constraints tend to be in better agreement with the predictions from the external shear model.
The main difference between Model~\texttt{III} and Model~\texttt{IV} is the value of the $\sigma_{0}$ predicted for the GGL lens. With the scaling relation and the values obtained for the standard galaxy, the GGLs lens is predicted to have $\sigma_{0}=114^{+26}_{-26}$\,km.s$^{-1}$ in Model~\texttt{IV} which differs from the value obtained with Model~\texttt{III}, $\sigma_{0}=179^{+3}_{-4}$\,km.s$^{-1}$.

\subsection{MACS~J0329}

The model of MACS\,J0329 (z$_{c}$~=~0.45) includes 177 cluster members plus the BCG and two cluster-scale halos for which the positions are shown in Fig.~\ref{fig:selection}. Following \citet{Zitrin2012}, the model is constrained by three multiple image systems (Table~\ref{tab:m0329mul}), systems \#1, \#2 and \#3. The redshift of system \#1 is fixed to the well-constrained photometric redshift $z_{s1}=6.18$. \citet{Zitrin2015} gives a spectroscopic redshift for system \#2, $z_{s2}=2.14$. The redshift of system \#3 is included as a free parameters in our model.

The GGL found in MACS\,J0329 is located South of the cluster. It is separated by 92\arcsec from the BCG (see Fig.~\ref{fig:selection}). As for RXJ2129-GGL1, we note that the multiple images are being sheared in a direction almost perpendicular to the direction of the cluster centre. Based on the spectroscopic redshift for the lens and the source, $z_{l}=0.3835$ and $z_{s}=1.112$ respectively (Sect.\ref{sec:specz}), the lens is a foreground galaxy and not a cluster member.

\begin{figure}
\centering
\includegraphics[width=\linewidth]{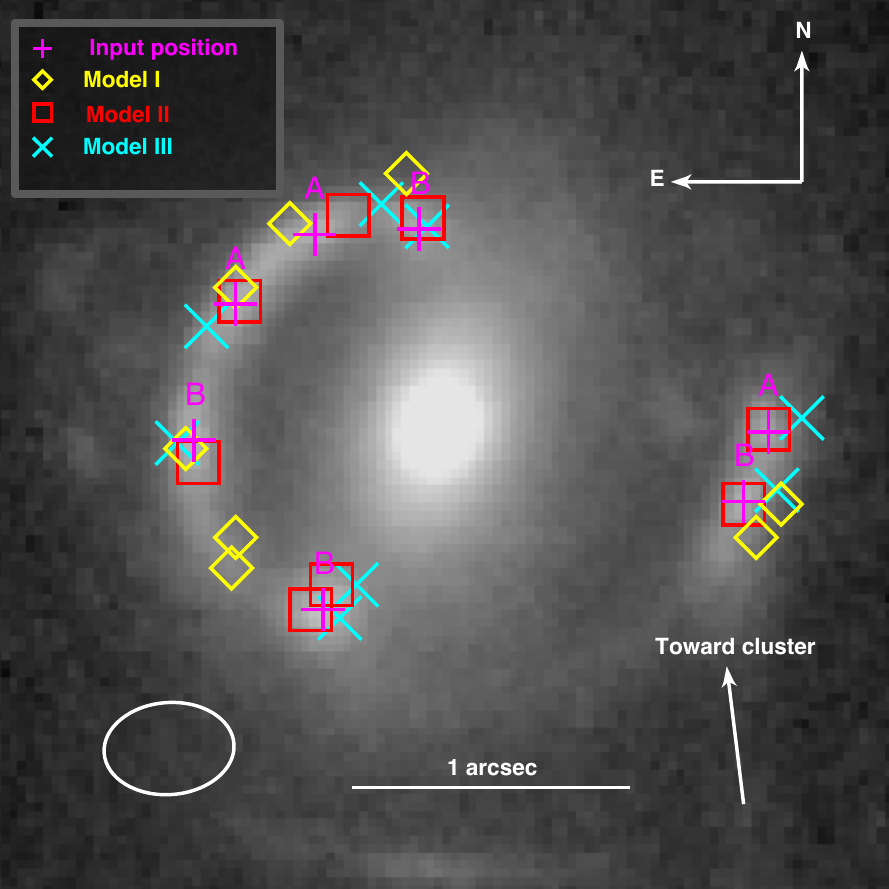}
\caption{Same as figure~\ref{fig:rxjpos} for MACS\,J0329-GGL1.}
\label{fig:m0329pos}
\end{figure}

Morphologically, the GGL system can be split into two different regions of similar colors which positions are listed in Table~\ref{tab:m0329mul} and shown Fig.~\ref{fig:m0329pos}. Each of them produces 4 multiple images, with A.4 and B.4 being coincident. We constrain the GGL using images A.1 to A.3 and B.1 to B.4. 
The best-fit model for this GGL is presented in Table~\ref{tab:m0329params}.

\subsubsection{Model \texttt{I}}

For the lens, the core radius is neglected and fixed to 0, and the $r_{cut}$ is arbitrarily set to a value of 50\,kpc.
The best-fit model predicts the position of the multiple images with an RMS of 0.20\arcsec (see table~\ref{tab:rms}). Figure~\ref{fig:m0329pos} shows the predicted positions of the multiple images with yellow diamonds. We see that the prediction reproduces the observed general shape of the system, but does not accurately recover the position of each multiple images.

\subsubsection{Model \texttt{II}}

Following the method described previously, we build a model that constrain the GGL lens parameters and the amplitude and orientation of a constant external shear at the redshift of the cluster. The addition of the shear brings more precision on the prediction of the multiple images as shown in Fig.~\ref{fig:m0329pos} (red boxes) and in Table~\ref{tab:rms} with an RMS of 0.07\arcsec. We note that the main arc and the counter-image in this system are unusually separated in the east-west direction, which is similar to the orientation of the shear as illustrated with the ellipse in the lower-left corner of Fig.~\ref{fig:m0329pos}. This ellipse shows that the shear seems to be oriented almost perpendicular to the cluster BCG direction. 

\begin{figure}
\centering
\includegraphics[width=\linewidth]{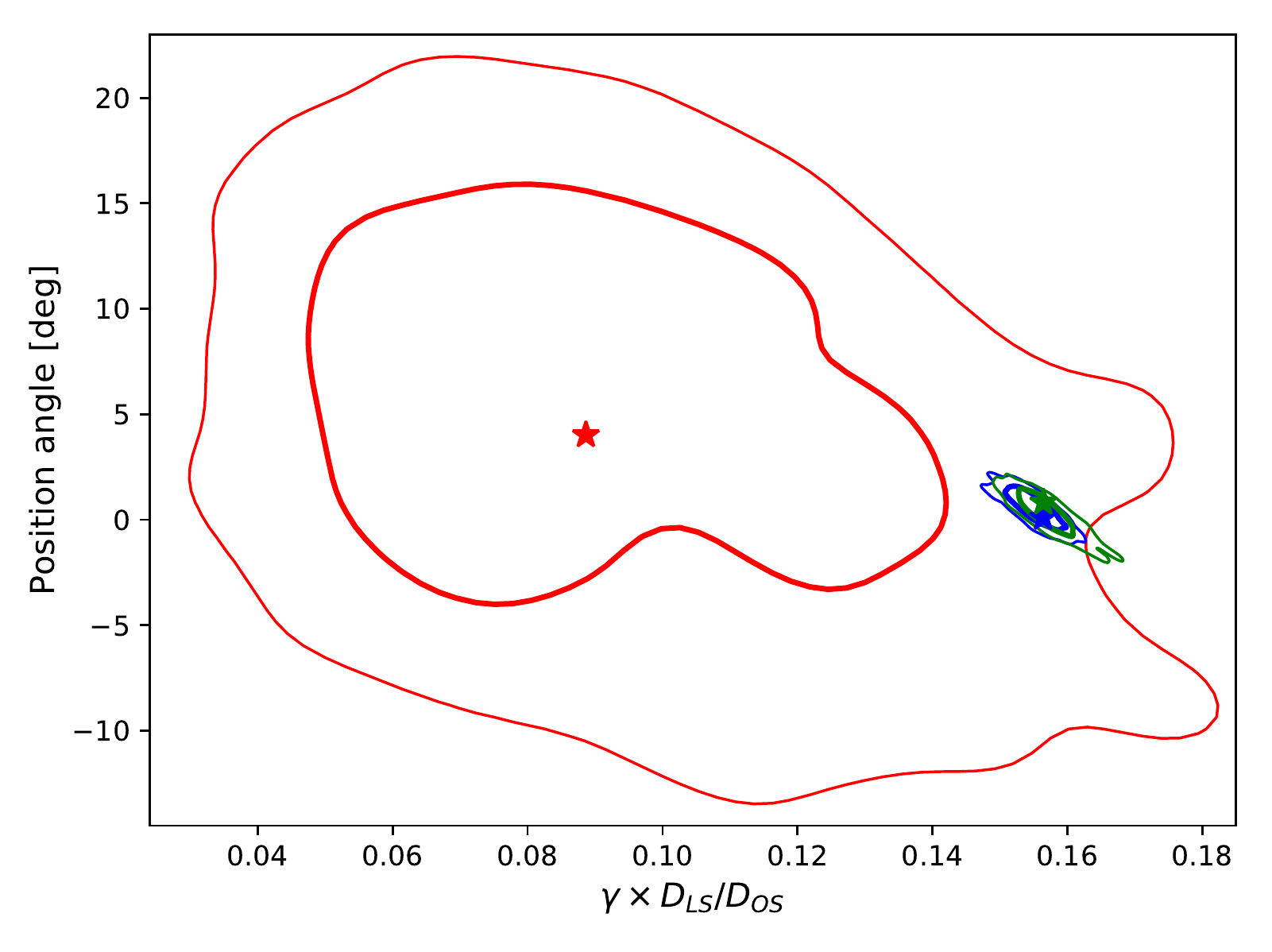}
\caption{Same as Fig.~\ref{fig:shearrxj} for MACS\,J0329 models.}
\label{fig:macs0329_shear}
\end{figure}

\subsubsection{Model \texttt{III}}

For this model, the two cluster-scale components were modelled by two dPIEs with a cut radius r$_{cut}$ of 1000\,kpc. \texttt{lenstool} optimises all the other parameters of the profile. The first halo position is centred on the BCG and allowed to vary within a $5\arcsec\times5\arcsec$ box. The second halo is allowed to move in a 30\arcsec $\times$50\arcsec\ area around its input position (RA: 52.4131055 ; Dec: -2.1914207). The BCG and the GGL lens are optimised as galaxy-scale dPIE potentials. Only their velocity dispersion is being optimised. The other parameters are fixed to the observed light distribution and typical values for galaxy and BCG potentials assuming they follow the scaling relation. 
The cluster members are being optimised following the Faber-Jackson scaling relation \citep[][]{FJ76} using the $F160W$-band as reference. The model is constrained by all the multiple images systems presented in Table~\ref{tab:m0329mul}.
The best-fit parameters are given in Table~\ref{tab:m0329params}. The RMS obtained is 0.10\arcsec\ compared to 0.07\arcsec\ for Model~\texttt{II}. The predicted multiple images are shown in Fig.~\ref{fig:m0329pos} as cyan crosses. The overall shape of the system is well recovered even if system \#A seems to be predicted with less precision than system \#B.

The best-fit model give a redshift z$_{s3}$~=~2.58$~\pm$~0.05 for system \#3. This value is in good agreement with the one derived by \citet{Zitrin2015}: $2.15<z<3.39$.

\subsubsection{Model \texttt{IV}}
The central galaxy of MACS\,J0329-GGL1 is not a cluster member. Thus, the GGL lens is not included in MACS\,J0329 cluster Model~\texttt{IV}. 
The resulting shear magnitude and orientation measurements are plotted in Fig.~\ref{fig:macs0329_shear}. Their values overlap with the ones from Model~\texttt{III} but are slightly more extended toward higher shear magnitude. Both of them remain within the 2$\sigma$ contours of Model~\texttt{II}.
Figure~\ref{fig:macs0329_shear} shows that this model of MACS\,J0329 tends to overestimate the amplitude of the shear at the location of the GGL. Also, the addition of the GGL in the model does not seem to constrain the shear at its particular location.  
The predicted redshift for system \#3 is z$_{s3}=2.59^{+0.06}_{-0.05}$, in good agreement with our previous results.

\subsection{MACS~J1149}

We used the MACS\,J1149 (z$_{c}=0.544$) model presented in \cite{Jauzac2016b}. This model combines 5 cluster-scale halos (see Fig.~\ref{fig:selection}) with 212 galaxy-scale haloes modelling cluster members. The model is constrained by 65 systems of multiple images.

\begin{figure}
\centering
\includegraphics[width=\linewidth]{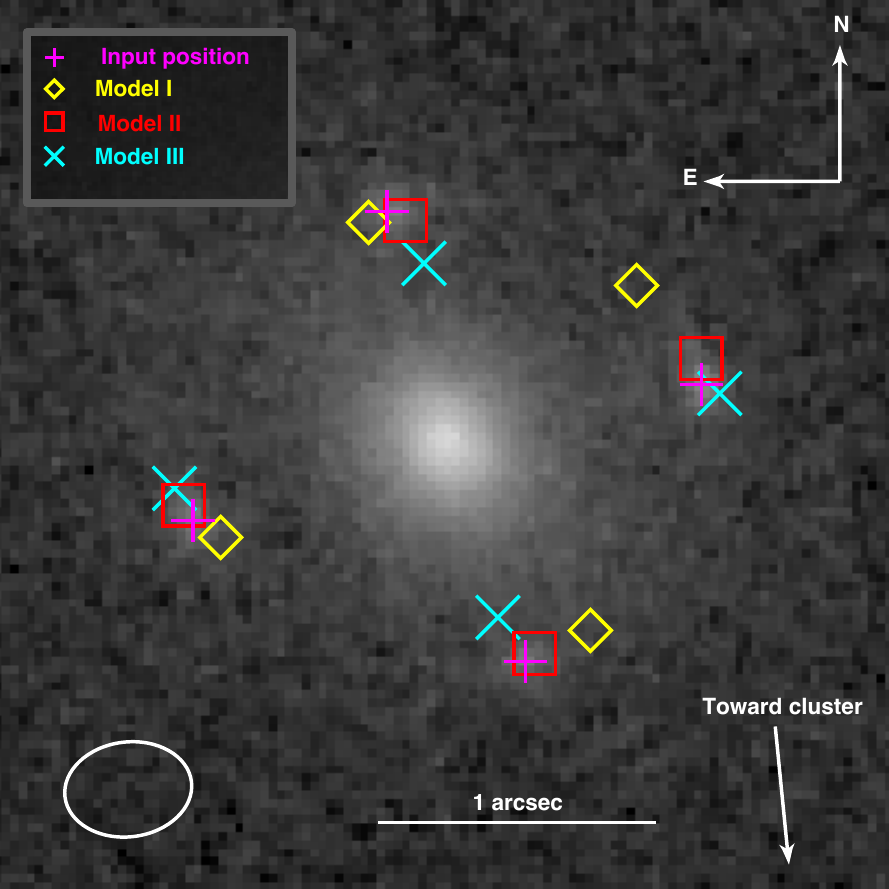}
\caption{Same as Fig.~\ref{fig:rxjpos}. for MACS\,J1149-GGL1}
\label{fig:m1149pos}
\end{figure}

MACS\,J1149-GGL1 is located North of the BCG at a distance of 137.9\arcsec\ (see Fig.~\ref{fig:selection} and Table~\ref{tab:Selected_GGL}). In Fig.~\ref{fig:selection}, the right panel shows the Einstein cross with its four images well separated from the lens. The lens galaxy has a measured spectroscopic redshift of $z_{l}=0.542$, compatible with the cluster redshift. The source has a measured spectroscopic redshift of $z_{s}=1.806$ (see Table~\ref{tab:Selected_GGL}).

The \citet{Jauzac2016b} model did not include the lens as one of the cluster member, thus we added it as a new galaxy potential. Since the lens does not have photometry in the $F814W$-band used in the scaling relations, we correct the measured $F775W$-magnitude to $F814W$ using the predicted colours for an elliptical galaxy at the cluster redshift (using the empirical template from \citet{Coleman}) and use that value ($m_{F814W}=20.11$) for the scaling relation.

\begin{figure}
\centering
\includegraphics[width=\linewidth]{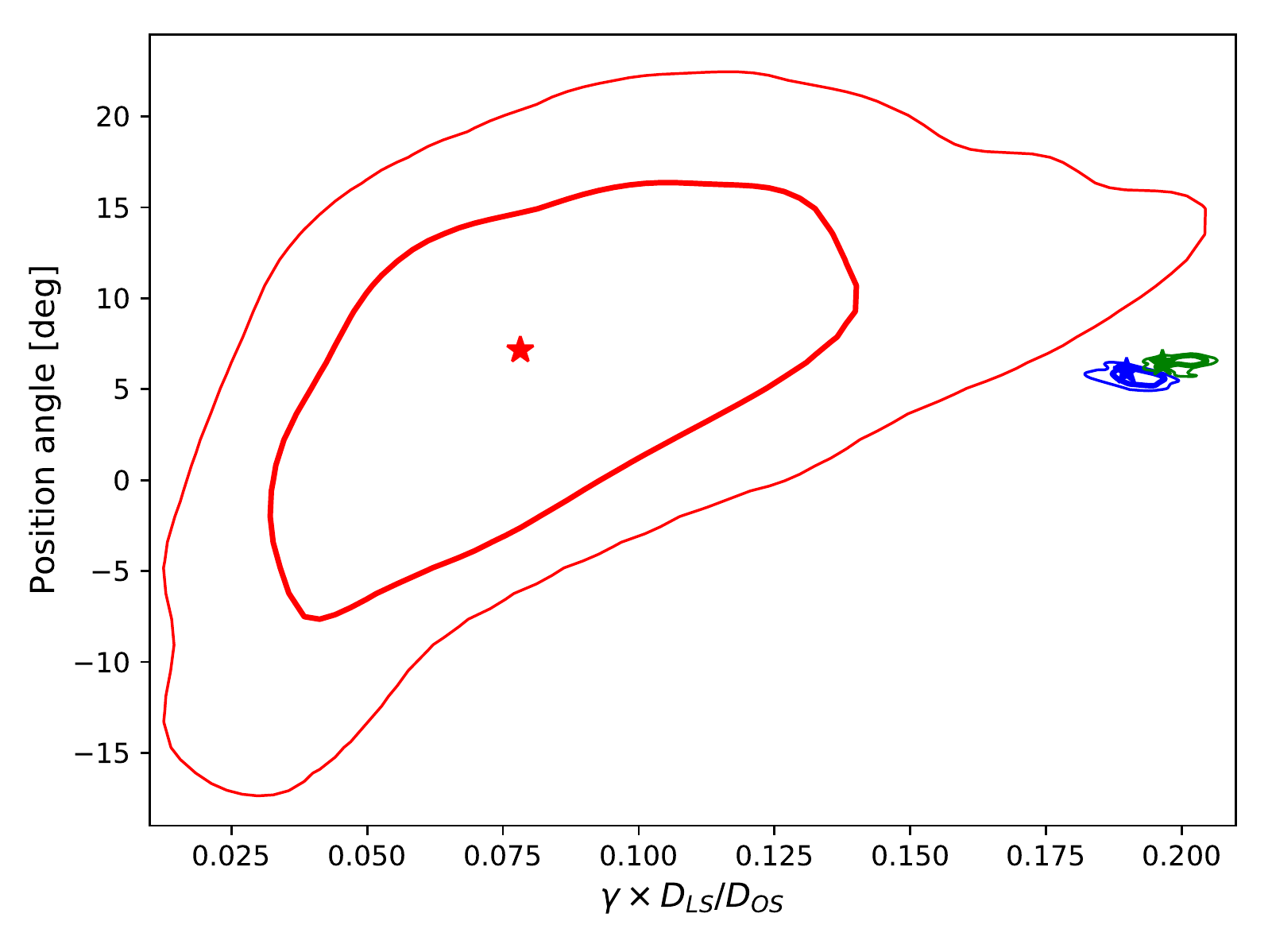}
\caption{Same as figure~\ref{fig:shearrxj} for MACS\,J1149 models.}
\label{fig:m1149shear}
\end{figure}

\subsubsection{Model \texttt{I}}

The GGL is modelled here as a galaxy-scale dPIE. The only parameter optimised is the velocity dispersion of the central galaxy. The geometrical parameters are being fixed to the ones from the light distribution while r$_{core}$ is set to 0 and r$_{cut}$ to 50\,kpc.
The predicted positions of the multiple images are presented in Fig.~\ref{fig:m1149pos}. They are aligned with the axes of the light distribution of the lens, but not the observed ones. The RMS is 0.26\arcsec (see Table.~\ref{tab:rms}).

\subsubsection{Model \texttt{II}}

The environment is modelled by a constant external shear constrained by the multiple images of the GGL together with the central lens. The resulting RMS is 0.07\arcsec. In Fig.~\ref{fig:m1149pos}, the predicted positions of the multiple images are shown by the red squares and are in good agreement with the observed ones. The ellipse in the bottom left corner of the figure represents the external shear and its orientation (perpendicular to the direction of the main cluster halo and the BCG).

\subsubsection{Model \texttt{III}}

This model is based on the work by \citet{Jauzac2016b} to which we add the potential of the GGL lens. The list of constraints is presented in Table~\ref{tab:m1149mul}. We include all the multiple image systems from the \citet{Jauzac2016b} model, but only use the central bulge of system \#1 as constraints and not all the star-forming regions of this spiral lensed galaxy.

The predicted positions of the multiple images can be seen on Fig.~\ref{fig:m1149pos} as cyan crosses. The East and West images are well predicted, while the North and South images are predicted closer to the lens than observed. This may be due to a more important shear than the measured one as shown in Fig.~\ref{fig:m1149shear}. 
The shear intensity is predicted higher in this model that in Model~\texttt{II}. There is still an improvement with respect to Model~\texttt{I} in predicting the multiple images positions, with an RMS of 0.17\arcsec\ (Table~\ref{tab:rms}).

Figures~\ref{fig:m1149pos} and~\ref{fig:m1149shear} show that the local shear magnitude of this model is overestimated by a factor 2.5 compared to the external shear model prediction. However its orientation is coherent with a difference smaller than 1.2 degrees compared the best predicted the external shear.

\subsubsection{Model \texttt{IV}}

This model is the same as Model \texttt{III}, without the multiple images of the GGL as constraints, and with the GGL lens optimised as a cluster member through the scaling relation. The measured shear is plotted in Fig.~\ref{fig:m1149shear}. It shows that the shear orientation is the same as the one measured in the two others models, but the shear magnitude is higher than the ones from Model \texttt{II} and Model \texttt{III}.


\section{Discussion}
\label{sec:discussion}

\subsection{GGLs parameters degeneracy}
\label{sec:degen}

In all our models, we fixed the value of the r$_{cut}$ parameter  in order to break its degeneracy with $\sigma_{0}$ according to \citet{Richard2010}. However, simple models optimising both parameters were made to check the status of the degeneracy using only GGLs constraints. For the three GGLs, the multiple images did not provide enough information to constrain r$_{cut}$. Yet, $\sigma_{0}$ is strongly degenerated for low r$_{cut}$ values but manage to be  constrained in those models due to its extremely low evolution with increasing r$_{cut}$ over 25~kpc. This indicates that for the typical values chosen for fixing r$_{cut}$ which are around 50~kpc, $\sigma_{0}$ is independent of this prior. Therefore, one can compare the results of optimisations of $\sigma_{0}$ without having to take in account the results on r$_{cut}$.

\subsection{Constraining the local shear with GGLs.}




For all three cases presented in this work, we find that including the detailed mass distribution of the cluster cores systematically improves the modelling of the GGL systems (Table~\ref{tab:rms}).
The RMS of the multiple images decreases in all cases by at least a factor of 1.5 with respect to the results obtained from models assuming a single galaxy lens alone. 

However we note that the best RMS are always achieved for models which include an external shear instead of a detailed cluster mass distribution. 
External shear models provide a measurement of the magnitude and orientation of the local shear due to the environment of the GGL without any knowledge of its nature. Our results suggest that the cluster itself is not the only shear source. 
One can argue about the robustness of a model that simple, and therefore the precision of the constraints the GGL provides on the local shear. For example, the knowledge of the source redshift can add some systematic uncertainties on the shear measurements. We can also test the values obtained against independent measurements coming from weak-lensing.


\begin{table}
\centering
\begin{tabular}{l c c c }
\hline
Cluster ID & Model \texttt{I} & Model \texttt{II} & Model \texttt{III} \\
\hline
MACS0329 & 0.20\arcsec\ & 0.07\arcsec\ & 0.10\arcsec\ \\
MACS1149 & 0.26\arcsec\ & 0.07\arcsec\ & 0.17\arcsec\ \\
RXJ2129 & 0.66\arcsec\ & 0.02\arcsec\ & 0.03\arcsec\ \\
\hline
\end{tabular}
\caption{RMS of the predicted positions of the multiple images of the GGLs systems with the different models, given in arcsec.}
\label{tab:rms}
\end{table}

\subsubsection{Impact of source redshift}
\label{sec:photozdiscuss}

\begin{figure}
\centering
\includegraphics[width=\linewidth]{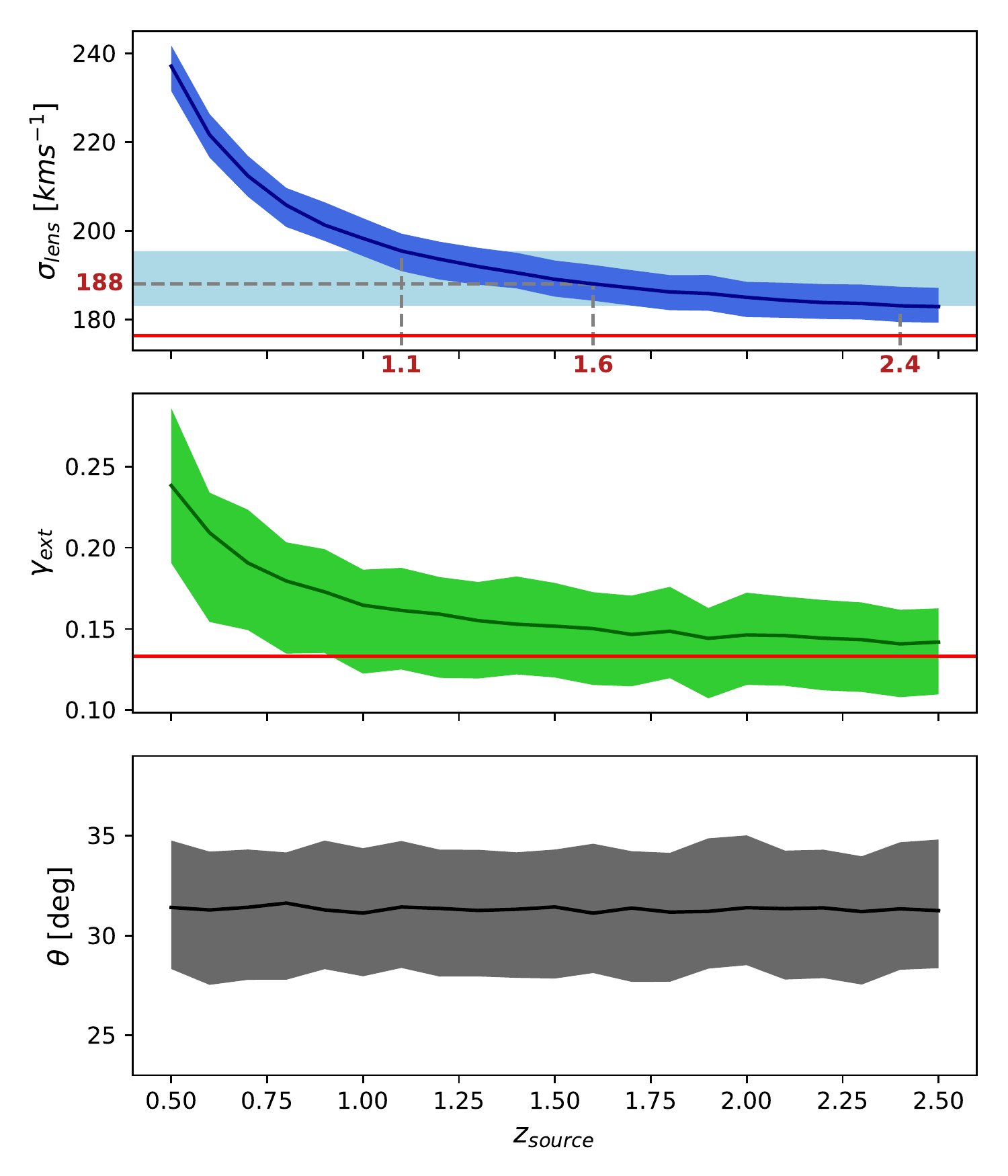} 
\caption{Velocity dispersion of the \emph{snail} lens, external shear amplitude and orientation \emph{versus} the redshift of the GGL source. The red lines show the values for $z_{s}=10$. In the upper panel, the grey lines show $z_{s}=1.1$, $z_{s}=1.6$, $z_{s}=2.4$, and $\sigma_{0}=188~km~s^{-1}$, the light blue area shows the variation on $\sigma_{0}$.}
\label{fig:zsnail}
\end{figure}

Among the three possible maxima of the photometric redshift probability distribution of the \emph{snail} (Sect.~\ref{sec:photoz}, Fig.\ref{fig:pdf}), we have so far assumed the middle peak $z=1.61$ for our models. 
Both the external shear and the velocity dispersion of the lens are degenerated with the source redshift, thus none of them can directly constrain the redshift. We build a series of models with external shear for different fixed source redshifts between $z=0.5$ and $z=2.5$, letting $\sigma_{0}$, the shear magnitude, $\gamma$, and its orientation, $\theta$, being optimised. The results of this test are presented in Fig.~\ref{fig:zsnail}. 
Under the assumption that the lens of the \emph{snail} follows the general scaling relation of cluster members (Sect.~\ref{sec:clustermodel}), its velocity dispersion should be $\sigma_{0} = 188$~km.s$^{-1}$. This indicates that a source redshift $z=1.6$ corresponds better to this assumption than $z=1.1$ or $z=2.4$. 

Figure~\ref{fig:zsnail} shows the evolution of the lens and the shear parameters as a function of the redshift. First, we note that the orientation of the shear is independent of the source redshift. Then, we observe that $\sigma_{0}$ and $\gamma$ have a strong evolution for redshift $z<1$. For redshift $z>1$, the evolution is slower, thus the variation on the values of $\sigma_{0}$ and $\gamma$ due to redshift uncertainties are less important. For a source at redshift $z=1.1$, the resulting velocity dispersion for the Snail lens would be $\sigma_{0}~=~195~\pm~5~km~s^{-1}$. This result varies of 3.7\% compared to the one presented in Table~\ref{tab:rxjparams}. The variation of $\gamma$ is 6.7\% from $\gamma=0.15^{+0.04}_{-0.03}$ for $z_{s}=1.61$ to $\gamma = 0.16^{+0.04}_{-0.03}$ for $z_{s}=1.1$. The variation on the results of $\gamma$ is smaller than the statistical errors from the models and the variation on $\sigma_{0}$ results have the same order of magnitude as the statistical errors of the models. Therefore, a photometric redshift seems precise enough to derive the properties of the lens and its environment in the case of a simple model.

\subsubsection{Comparison with weak-lensing constraints}
\label{sec:wl}

Weak-lensing is the usual measurement to be used to estimate the shear produced by the direct environment. By measuring the shape of the background sources as observed in the cluster we obtain an independent estimation of the shear signal at large radii from the core (i.e. outside the strong-lensing region). 

Following the methodology described in \cite{Jauzac2012,Jauzac2015a}, we construct the background galaxy catalogues using the HST data. We only give a brief description and refer the reader to the former papers for more details. The detection of sources is done using \textsc{sextractor} \citep{Bertin1996} in the $F814W$-band, and the galaxy shapes are measured using the RRG method \citep{Rhodes2000}. RRG was developed for measurements on HST/ACS observations and therefore includes corrections of the point-spread function (PSF). One of the careful steps in the build-up of the weak-lensing catalogue is the removal of the foreground and cluster galaxies that would otherwise dilute the shear signal. To counteract this problem, as we do not have a redshift for all sources, we identify the regions populated by these different galaxy populations in the colour-colour space $mag_{F435W}$-$mag_{F606W}$-$mag_{F814W}$, and exclude them from our final catalogue. This colour-colour selection is calibrated using the publicly available photometric redshifts from the CLASH collaboration\citep{Postman2012}. We further apply standard lensing cuts: (1) on the size of the galaxies to remove galaxies with a size close to the one of the PSF ($>0.13\arcsec$), and (2) on the detection limit of the sources with a signal-to-noise ratio (SNR) greater than 4.5.
Our final catalogue contains 385 galaxies, therefore a density of background sources of $\sim$50\,gal.arcmin$^{-2}$.

From this weak-lensing catalogue we can then measure both the tangential and radial shear profiles for RXJ2129, $\gamma_{t}$ and $\gamma_{x}$ respectively, using the following inversion relations:

\begin{equation}
\gamma_{t} = -(\gamma_{1}\times cos(2\alpha) + \gamma_{2}\times sin(2\alpha)) \ ,
\end{equation}
and
\begin{equation}
\gamma_{x} = -\gamma_{1}\times sin(2\alpha) + \gamma_{2}\times cos(2\alpha) \ ,
\end{equation}

where $\gamma_{1}$ and $\gamma_{2}$ are provided by RRG and $\alpha$ is the position angle between the vector pointing in the decreasing RA direction (West) and the vector connecting the BCG to the background source. 
%
%
As the redshift of all weak-lensing galaxies is not known, we need to assume a background redshift distribution. For this purpose we make use of the \emph{Hubble Frontier Field} Abell\,2744 photometric redshift catalogue provided as part of the HFF-DeepSpace project \citep{Shipley2018}. We only consider the distribution of sources at a redshift higher than the clusters RXJ2129 and Abell\,2744, i.e. $z>0.4$, and with a photometric redshift error better than 10\%. We further apply a magnitude cut, $m_{F814W}<25.5$, in order to match the depth of the RXJ2129 images. Random redshifts are drawn from this distribution and assigned to our catalogue sources. The average $\gamma_{t}$ and $\gamma_{x}$ are then calculated in annuli of 20\arcsec centred on the BCG. This process is repeated 100 times, and the final values considered here are the means and their respective standard deviations of these 100 realizations.

\begin{figure}
\centering
\includegraphics[width=\linewidth]{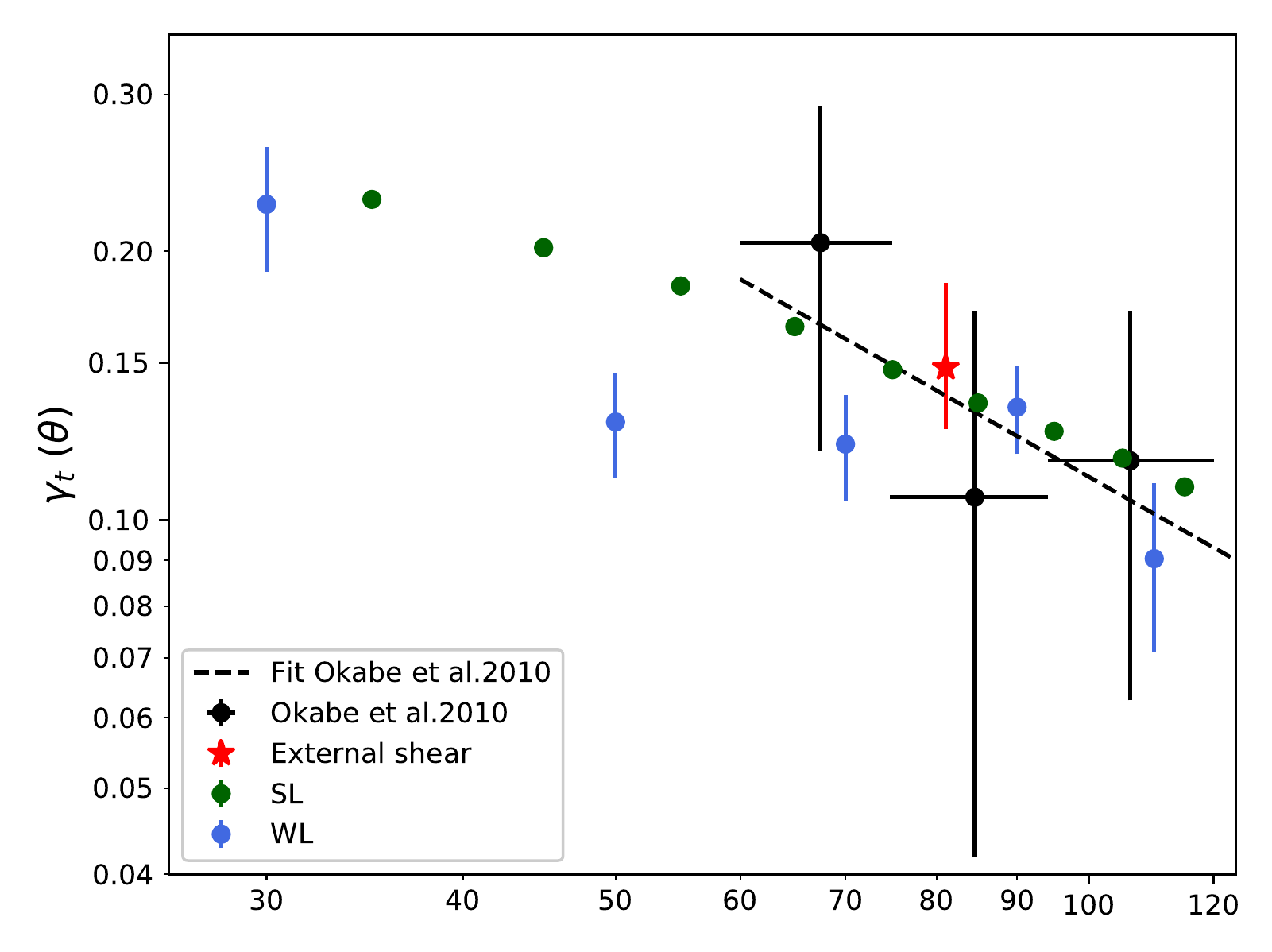} \\
\includegraphics[width=\linewidth]{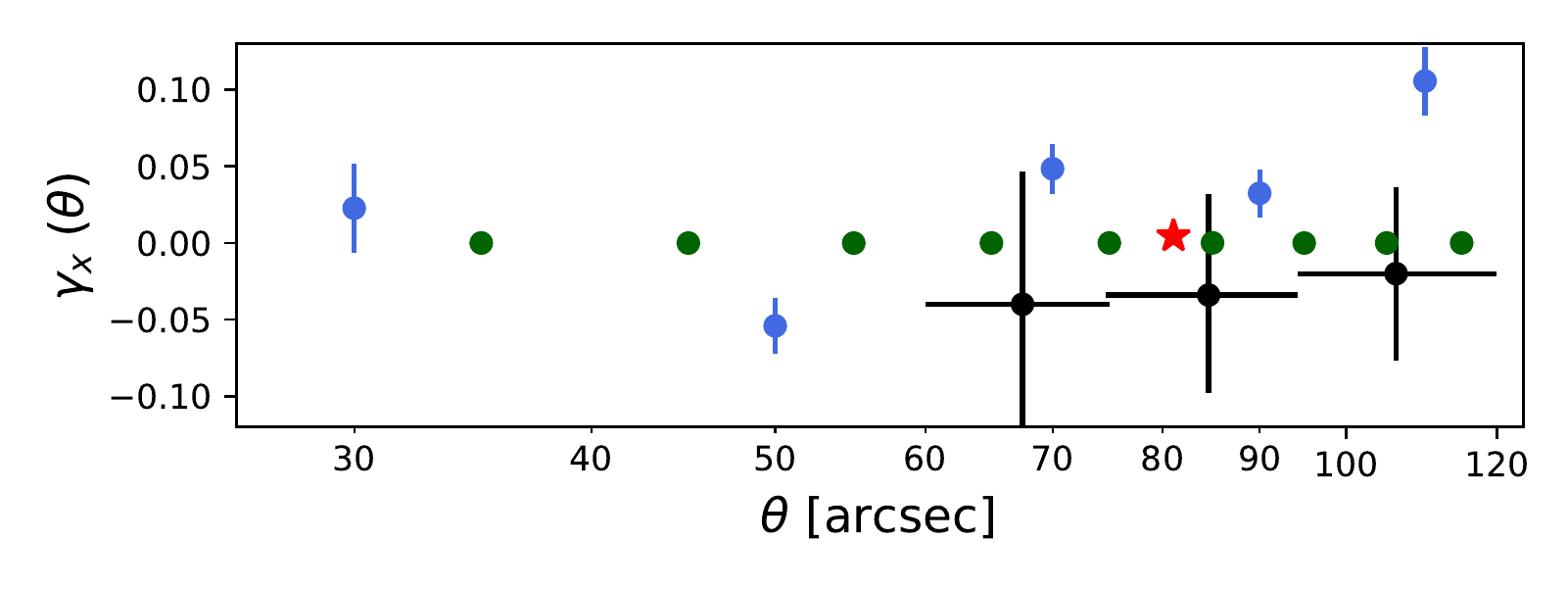}
\caption{Shear component profile (tangential $\gamma_{t}$ and radial $\gamma_{x}$) as a function of radius from BCG. In blue, the shear derived from the CLASH data weak lensing analysis. In green the shear predicted by the RXJ2129 complete cluster model. In black, the shear measurement and the predicted shear with the SIS model from \citet{Okabe2010}. The shear estimated with the external shear model is represented with the red star.}
\label{fig:wl}
\end{figure}

Figure~\ref{fig:wl} shows the comparison of the tangential ($\gamma_{t}$) and radial ($\gamma_{x}$) shear profiles as a function of the radius from the BCG obtained with different measurement methods: (1) the weak-lensing analysis from high-resolution HST images (blue filled circles), and (2) the predicted external shear from the strong lensing model of the cluster core (green filled circles). 
These profiles are also compared to the shear profile measured by \citet[][; black filled circles]{Okabe2010} and their Single Isothermal Sphere (SIS) fit (black line). 
The external shear value from Model~\texttt{II} is highlighted by the red star.

At the location of the \emph{Snail} (a region comprised within $60\arcsec$ and 120$\arcsec$ from the cluster BCG), we observe an excellent agreement between the weak-lensing shear measured in this paper, the strong lensing extrapolation, the measurements from \citet{Okabe2010}, and the predicted shear value from the external shear model (Model~\texttt{II}). 

Both direct weak-lensing measurements show a really good agreement. The ground-based values from \cite{Okabe2010} have larger error bars due to the lower background galaxy density, $\sim$30\,gal.arcmin$^{-2}$, compared to our HST measurement. 

We further compare our HST weak-lensing measurement with the predicted external shear of Model~\texttt{II}. One can see that the predicted external shear is similar to the HST weak-lensing shear, including its error estimate. This agreement reveals the potential for galaxy-galaxy lensing to locally probe the shear profile in the outskirts of clusters. The annulus around the GGL radius encompasses $\sim$100 weak-lensing background galaxies ($80\arcsec<R<100\arcsec$) and thus have a local source density of $\sim$35~gal.arcmin$^{-2}$. That means a single GGL event in an area of $\sim$9\,arcsec$^{2}$ provides a shear measurement equivalent to a standard HST weak-lensing analysis over an area of $\sim$3\,arcmin$^{2}$. However, this is only true when the studied cluster is being relaxed, i.e. no substructures in its outskirts.
In our sample of GGLs, only RXJ2129-GGL1 is observed in a relatively relaxed cluster. This is why we used it to show the strength of GGL local shear measurements.
%
%



\subsection{Simple constraints on the cluster based on external shear}
\label{sec:sheardiscission}


\citet{HTu2008} showed that some partial information about the cluster mass distribution can be retrieved purely based on GGL analysis, as they derive the position of the centre of Abell 1689 cluster with 3 GGLs. Here we test whether we can blindly retrieve the directions of the clusters from the GGLs positions and estimates of their central velocity dispersion from the external shear models, under the assumption that this shear is dominated by the presence of the cluster.

We can give the direction of the centre of the cluster for RXJ2129 using the orientation of the shear that is supposed to be perpendicular to the direction toward the cluster. The values computed for all the cluster in our study are in Table~\ref{tab:SIS}. The measured angle from the GGL to the BCG is $-57.0\pm0.4$ degrees, and the angle given by the external shear is $-58.5^{+3.3}_{-3.2}$ degrees. We have a good agreement on this value.

Assuming that the cluster modelled by a SIS, we can also compute the velocity dispersion of the cluster from the shear magnitude $\gamma$ \citep{Dye2007}. The relation between $\sigma_{0}$ and $\gamma$ is:
\begin{equation*}
\sigma_{0} = \sqrt[]{\frac{\gamma c^{2} R}{2 \pi D_{LS}/D_{OS}}}
\end{equation*}
where R is the distance between the GGL and the cluster  and $D_{LS}/D_{OS}$ is the ratio of the angular diameter distances between the lens and the source, and between the observer and the source. In the case of RXJ2129 we obtain $\sigma_{RXJ2129} = 912^{+105}_{-70} km~s^{-1}$. This result is matching the results of the one of the complete cluster model (see Table~\ref{tab:rxjparams}) even if we only assume here the contribution of the cluster clump of DM. We can note also that our result is also in agreement with the $\sigma_{SIS}$ of \citet{Okabe2010}.

As seen in the section~\ref{sec:photozdiscuss}, for RXJ2129, the photometric redshift of the Snail source increases our uncertainty on the external shear. The result of the process using the shear compatible with a source at $z_{s}=1.1$ is $\sigma=952_{-81}^{+102}km~s^{-1}$ thus a 4.3\% variation to the previous result. This value is still consistent  with the complete model of the cluster within error bars, and once again the variation is less significant than the error on the value. Finally for RXJ2129, as seen in Fig.~\ref{fig:zsnail}, the result for the direction of the cluster remains unchanged with the change of redshift.

\begin{table}
\centering
\begin{tabular}{l l l l }
\hline
ID & $\theta_{obs}$ & $\theta_{Mod}$ & $\sigma_{SIS}$ \\
 & [deg] & [deg] & km/s \\
\hline
RXJ2129 & -57.0$\pm$0.36 & -58.5$^{+3.3}_{-3.2}$ & $912^{+105}_{-70}$ \\
MACS0329 & 97.0$\pm$0.32 & 94.0$^{+11.3}_{-2.1}$ & $1041^{+222}_{-131}$ \\
MACS1149 & -83.3$\pm$0.21 & -83.2$^{+10.9}_{-4.7}$ & $1198^{+349}_{-172}$ \\
\hline
\end{tabular}
\caption{Table of the observed direction angle from the GGLs toward the BCGs  and the computed angle from the perpendicular direction of the shear from Models \texttt{II}. The last column is the velocity dispersion of an SIS at the position of the BCG derived from the shear magnitude. The error of $\theta_{obs}$ is the result to the propagation  assuming a positional error of 0.5\arcsec for the BCG and the GGLs.} 
\label{tab:SIS}
\end{table}

The same procedure was applied for the two other clusters and all the results can be found in Table~\ref{tab:SIS}. We can see that the predicted orientation of the shear is a good indicator of the position of the cluster centre. Comparing the velocity dispersions of the SIS to the ones predicted by Model  \texttt{III} (see Tables~\ref{tab:rxjparams},\ref{tab:m1149params} and \ref{tab:m0329params}) we note that only the only cluster with a good agreement is RXJ2129. This can be explained by the simplicity of the cluster structure, only one cluster halo of DM, and thus the absence of substructures in its surroundings. Also, we only assumed here the contribution of the cluster but not the one from the BCG and the cluster members. That could explain the systematic higher value of $\sigma_{0}$ for all of the three clusters. In any case, this method provide a blind estimate of the cluster velocity dispersion without the need for constraints by multiple images near its core.

\subsection{Combining GGLs with cluster core models}




The strong lensing constraints of the GGLs allow to measure locally the influence of the cluster at large radii. But this influence is only a second-order effect, as the clusters enhance the lensing power of the single galaxies and produce a shear. Figures~\ref{fig:shearrxj}, \ref{fig:macs0329_shear} and \ref{fig:m1149shear} show that the GGLs constraints only have a small influence on the cluster core models. We can see that the contours of the complete models of the clusters tend to get closer to the results of the external shear models when the GGLs constraints are taken in account, but the shear is not perfectly reproduced, leading to a higher RMS in the prediction of the multiple images than models with external shear (see Table~\ref{tab:rms}). This lack of influence can be the fact of the GGLs constraints being only one more system of multiple images among others that are closer to the core, thus having more influence. The clusters parametric models might be too constrained by those multiple images in the cores to reproduce correctly both the core and the outskirts structures. New parameters bringing new degrees of freedom, especially in the outskirts could be a solution as long as they do not lead to an over-fit of the model. One other explanation of the difference between external shear models and complete cluster models results would be that the influence of the cluster is only a part of the environment shear. Fig.~\ref{fig:selection} shows that for MACS0329 and MACS1149, the GGLs are at the edge of the ACS data. The environment influence might not be completely accounted for, thus explaining the small difference made by the addition of the GGLs constraints in Fig~\ref{fig:macs0329_shear} and \ref{fig:m1149shear}. For RXJ2129, the GGLs is closer to the BCG than in the two other cases, thus its environment is better known and the shear prediction of the cluster model (Fig~\ref{fig:shearrxj}) seems more affected by the GGL constraints, supporting this solution.


For GGLs for which the lens is part of the cluster, the multiple images directly constrain the massive cluster members. If the spectroscopic redshift of the source is known, we can determine if the galaxy lies on top of the scaling relation or not by having an independent measurement of its parameters.
For MACS1149-GGL1, we know the redshift and thus we can compare the  values of the model with the expected scaling relations described in Sect.\ref{sec:clustermodel} using $\sigma_{0}^{*}=158$ km~s$^{-1}$ from \citet{Bernardi2003}. The expected value is $\sigma_{0}=178^{+31}_{-30}~km~s^{-1}$ which is in agreement with the model values in Table~\ref{tab:m1149params}, mostly the one of the most complete cluster model. The two other results are closer to the upper limit value because those models do not take in account the impact of the cluster convergence boosting the lensing power of the galaxy, therefore leading to an overestimation of the velocity dispersion of the lens. However the value of $\sigma^{*}_{0}$ of the complete cluster model is $\sim$~40\% higher than the one optimised in the cluster model. This could indicate that the standard galaxy is not constrained well enough, as the cluster model not including the GGL constraints provide different values. We find a similar problem in the cluster model of RXJ2129 where $\sigma^{*}_{0}=93\pm16~km~s^{-1}$. If the GGL lens follows the scaling relation as we assumed it, the velocity dispersion of the lens would be $\sigma=114~km~s^{-1}$ which is far too low according to Fig.~\ref{fig:zsnail} to produce multiple images as we observe them even for a source with $z_{s}=10$. Even with the boost of the cluster, a complete model constraining the GGL parameters with a source with z$_{s}=10$ leads to a $\sigma_{0}$=166.8~$km~s^{-1}$. There is a $\sim$~30\% variation with this value and the one derived from the scaled standard galaxies in the models presented in Table~\ref{tab:rxjparams}. Either the standard galaxy parameters are not well constrained or our assumption about the GGL lens is wrong. Having a spectroscopic redshift for the GGL source would provide a way to constrain $\sigma_{0}$ independently of the scaling relation and would allow us to test the consistency of the results. Then assuming that the galaxy follows this relation, we could constrain better the standard galaxy parameters directly using the locations of multiple images in the GGLs. For this reason, spectroscopic follow-up of the 24 GGLs presented in Fig.~\ref{fig:mosaique} and Table~\ref{tab:GGLcat} would improve greatly the model constraints for all those clusters.
%
%

\section{Summary and conclusions}


We visually inspect the full Hubble field-of-view of the 25 observed clusters from the CLASH survey in order to locate GGL events in the outskirts of those clusters. We find a selection of 24 candidate GGLs (some already known),  and study in detail three of them presenting the following characteristics: a single lens, at least 4 distinct multiple images, and a separation from the BCG larger than 80\arcsec. For each of those GGLs and their associated cluster, we produce 4 parametric models of the DM distribution to study the influence of the cluster on the GGL modelling and the influence on the GGL on the cluster models.

Through those models, we show that the modelling of the GGLs cannot be done properly without taking into account its environment. This can be achieved through a complete model of the neighbour structures or even with a simple parametrisation of their effects like an external shear. 

A photometric redshift is accurate enough to properly estimate the strength of the shear as the uncertainties bring a variation that is smaller to the statistical errors on the measurement. The orientation of the shear is always well estimated as it is redshift independent. The measurement of the local external shear has a similar quality as independent measurements of the shear through weak lensing.

The constrained local shear magnitude and orientation are precise enough to properly derive the direction toward the cluster core, and its central velocity dispersion assuming a SIS distribution of the DM halo when the cluster structure is simple. For more complex clusters, the velocity dispersion of the central clump is overestimated. Therefore the strong lensing constraints of the GGLs allow an independent estimate or provide an upper-limit to the properties of a neighbour cluster without the need of multiple images in the core to constrain it.

When combined with a complete cluster strong-lensing model the first-order effect of the GGL constraints is to constrain with precision the DM halo of the lens galaxy. However, they bring only a little information to the parameters of the core as its influence is a second-order effect. Therefore the complete cluster models do not reproduce the GGLs multiple images as well as the external shear models do. This can be the sign of the parametric models not having enough freedom in the outskirts to constrain the DM distribution or that our knowledge of the environment is not complete enough as the GGLs lie at the edge of the ACS fields.

In the case of GGL lenses that are also cluster members, there are inconsistencies between the derived scaling relations and the GGL lens properties. The knowledge of the spectroscopic redshifts of the sources could allow to study the link between the massive cluster members in the outskirts and the scaling relation. 

A spectroscopic follow-up of the GGLs presented in this work would confirm their nature as GGLs, and bring independent estimates on the cluster mass profiles at large radii. For lenses located in the cluster, it could also bring constraints on the scaling relations assumed in galaxy cluster models.

\section*{Acknowledgements}

GD, JR, JM and BC  acknowledge support from the ERC starting grant 336736-CALENDS. GD also acknowledges a grant from the Swiss National Science Foundation. MJ was supported by the Science and Technology Facilities Council [grant numbers ST/L00075X/1 and ST/P000541/1]. We acknowledge fruitful discussions with Jean-Paul Kneib, Eric Jullo and Marceau Limousin. 
Based on observations made with ESO Telescopes at the La Silla Paranal Observatory under programme ID 186.A-0798. Based on observations obtained with the NASA/ESA Hubble Space Telescope, retrieved from the Mikulski Archive for Space Telescopes (MAST) at the Space Telescope Science Institute (STScI). STScI is operated by the Association of Universities for Research in Astronomy, Inc. under NASA contract NAS 5-26555. Also based on data obtained at the W.M. Keck Observatory, which is operated as a scientific partnership among the California Institute of Technology, the University of California and the National Aeronautics and Space Administration. The Observatory was made possible by the generous financial support of the W.M. Keck Foundation. The authors wish to recognise and acknowledge the very significant cultural role and reverence that the summit of Mauna Kea has always had within the indigenous Hawaiian community.  We are most fortunate to have the opportunity to conduct observations from this mountain. 
This work is based on data and catalog products from HFF-DeepSpace, funded by the National Science Foundation and Space Telescope Science Institute (operated by the Association of Universities for Research in Astronomy, Inc., under NASA contract NAS5-26555).

\bibliographystyle{mnras}
\bibliography{reference}







\appendix

\section{Galaxy-galaxy lensing in CLASH}
\label{appendixA}
We list here for reference all the GGL systems we have found from visual inspection of all the CLASH HST images.

\begin{table*}
\begin{tabular}{l c r r  l }
\hline
ID & Image & $\alpha$ & $\delta$ & Previous reference \\
\hline
A209-GGL1 & A & 22.9577568 & -13.6032558 & \\
A209-GGL2 & B & 22.9648793 & -13.6363138 & \\
A383-GGL1 & C & 42.0113589 & -3.5480288 & \\
MACS0429-GGL1 & D & 67.4020771 & -2.8713932 & \\
MACS0429-GGL2 & E & 67.3892478 & -2.8741192 & \\
MACS0329-GGL1 * & F & 52.4201304 & -2.2216321 & \\
MACS0416-GGL1 & G & 64.0340808 & -24.0667448 & ID14 from \citet{Vanzella2017}\\
MACS0416-GGL2 & H & 64.0284705 & -24.085668 & \\
MACS0416-GGL3 & I & 64.0170899 & -24.0895541 & "Dragon Kick" from \cite{DragonKick} \\
MACS0717-GGL1 & J & 109.3786176 & 37.77722736 & \\
MACS0744-GGL1 & K & 116.2121685 & 39.4598681 & \\
MACS1115-GGL1  & L & 168.9562589 & 1.4974098 & \\
MACS1149-GGL1 *& M & 177.4028221 & 22.4366292 & \\
MACS1149-GGL2 & N & 177.4116004 & 22.4296659 & \\
MACS1149-GGL3 & O & 177.403888 & 22.4266297 & A6 system from \cite{Smith2009} \\
MACS1149-GGL4 & P & 177.3931348 & 22.4113364 & A5 system from \cite{Smith2009} \\
RXJ1347-GGL1 & Q & 206.8960322 & -11.7536032 & \\
RXJ1347-GGL2 & R & 206.865999 & -11.7649203 & F system from \cite{Bradac2008} \\
RXJ1347-GGL3 & S & 206.8725903 & -11.7673974 & G system from \cite{Bradac2008} \\
MS2137-GGL1 & T & 325.0615233 & -23.6511738 & \\
RXJ2129-GGL1 * & U & 322.4287798 & 0.1080707 & \\
SMACS2248-GGL1 & V & 342.2156577 & -44.5183953 & \\
SMACS2248-GGL2 & W & 342.1557424 & -44.5459123 & \\
SMACS2248-GGL3 & X & 342.1633643 & -44.5297236 & \\
\hline
\end{tabular}
\caption{Catalogue of the GGL found in the CLASH data. When relevant, we provide the name used in previous works mentioning the same systems. The GGL studied in this work are pointed out with *}
\label{tab:GGLcat}
\end{table*}

\begin{figure*}
\includegraphics[width=\linewidth]{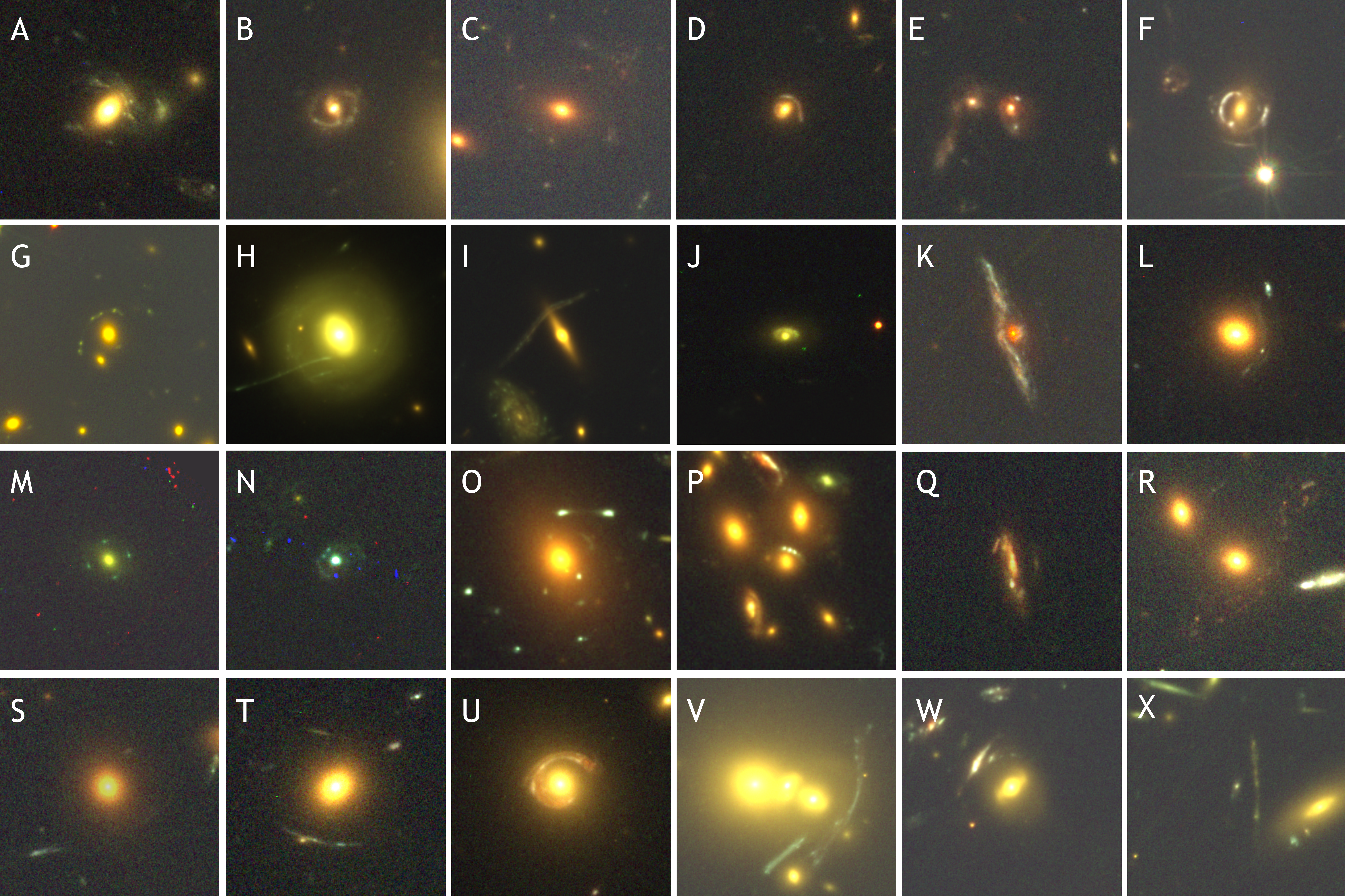} 
\caption{GGLs found in the CLASH images. All the pictures are 10\arcsec\ across. J B(F435W+F475W), G(F555W+F606W), R(F775W+F814W+F850LP). G, H, I, J, O, P, V, W and X, images frontier fields: R(F814W), G(F606W), B(F435W).}
\label{fig:mosaique}
\end{figure*}

\section{Multiple images systems in the clusters}
\label{appendixB}

\begin{table*}
\centering
\begin{tabular}{lllll}
\hline
ID & $\alpha$ & $\delta$ & $z_{\rm prior}$ & $z_{\rm model}$ \\
   & [deg] & [deg] & & \\
\hline 
 A.1 & 21:29:42.85 & 00:06:30.27 & 1.61 & \\
 A.2 & 21:29:42.96 & 00:06:29.99 & 1.61 & \\
 A.3 & 21:29:42.99 & 00:06:29.06 & 1.61 & \\
 A.4 & 21:29:42.88 & 00:06:28.45 & 1.61 & \\
 1.1 & 21:29:40.89 & 00:05:17.95 & 1.522 & \\
 1.2 & 21:29:40.84 & 00:05:23.15 & 1.522 & \\
 1.3 & 21:29:40.31 & 00:05:35.76 & 1.522 & \\
 3.1 & 21:29:40.44 & 00:05:07.68 & [0.2-2.0] & 1.49$^{+0.17}_{-0.09}$\\
 3.2 & 21:29:40.24 & 00:05:24.97 &  & \\
 3.3 & 21:29:39.77 & 00:05:31.99 &  & \\
 5.1 & 21:29:39.98 & 00:05:15.87 & [0.2-2.0] & 0.78$^{+0.05}_{-0.03}$\\
 5.2 & 21:29:39.90 & 00:05:17.17 &  & \\
\hline
\end{tabular}
\caption{Multiple images used in the RXJ2129 models. The results on the redshifts estimation are the ones of Model \texttt{III}.}
\label{tab:rxjmul}
\end{table*}

\begin{table*}
\centering
\begin{tabular}{llllllllll}
\hline
Potential & $\Delta\alpha$ & $\Delta\delta$ & $e$ & $\theta$ & r$_{\rm core}$ & r$_{\rm cut}$ & $\sigma_{0}$ & $\gamma$ \\
   & [arcsec] & [arcsec] & & [deg] & kpc & kpc & km/s \\
\hline 
\multicolumn{8}{l}{Model \texttt{I}} \\
GGL & $[-44.2]$ & $[ 68.0]$ & $[0.11]$ & $[-50.6]$ & $[0]$ & $[64]$ & $222^{+1}_{-1}$ \\\hline
\hline 
\multicolumn{8}{l}{Model \texttt{II}} \\
GGL & $[-44.2]$ & $[ 68.0]$ & $[0.11]$ & $[-50.6]$ & $[0]$ & $[64]$ & $188^{+4}_{-3}$ \\
Ext Shear &  &  &  & $ 31.5^{+  3.3}_{ -3.2}$ &  &  &  & 0.15$^{+0.04}_{-0.03}$\\
\hline   
\hline
\hline 
\multicolumn{8}{l}{Model \texttt{III}} \\
GGL & $[-44.2]$ & $[ 68.0]$ & $[0.11]$ & $[-50.6]$ & $[0]$ & $[64]$ & $179^{+3}_{-4}$ \\
DM1 & $  1.2^{+  1.4}_{ -1.2}$ & $ -1.0^{+  0.8}_{ -0.4}$ & $ 0.59^{+ 0.05}_{-0.06}$ & $-21.8^{+  0.3}_{ -0.4}$ & $49^{+8}_{-6}$ & $[1000]$ & $852^{+49}_{-27}$ \\
BCG & $[  0.0]$ & $[  0.0]$ & $[0.49]$ & $[-35.4]$ & $[0]$ & $[90]$ & $220^{+17}_{-20}$ \\
L$^{*}$ galaxy &  & & & & $[0.15]$ & $[45]$ & $93^{+16}_{-16}$\\
\hline
\multicolumn{8}{l}{Model \texttt{IV}} \\
DM1 & $  0.9^{+  1.3}_{ -1.1}$ & $ -0.8^{+  0.5}_{ -0.6}$ & $ 0.61^{+ 0.05}_{-0.07}$ & $-21.6^{+  0.4}_{ -0.4}$ & $44^{+5}_{-4}$ & $[1000]$ & $824^{+25}_{-27}$ \\
BCG & $[  0.0]$ & $[  0.0]$ & $[0.49]$ & $[-35.4]$ & $[0]$ & $[90]$ & $222^{+23}_{-19}$ \\
L$^{*}$ galaxy &  & & & & $[0.15]$ & $[45]$ & $96^{+22}_{-22}$\\
\hline
\end{tabular}
\caption{Parameters for the RXJ2129 models}
\label{tab:rxjparams}
\end{table*}

\begin{table*}
\centering
\begin{tabular}{lllll}
\hline
ID & $\alpha$ & $\delta$ & $z_{\rm prior}$ & $z_{\rm model}$ \\
   & [deg] & [deg] & & \\
\hline 
 A.1 & 11:49:36.61 & 22:26:12.08 & 1.806 & \\
 A.2 & 11:49:36.69 & 22:26:12.70 & 1.806 & \\
 A.3 & 11:49:36.74 & 22:26:11.59 & 1.806 & \\
 A.4 & 11:49:36.66 & 22:26:11.08 & 1.806 & \\
 1.1 & 11:49:35.28 & 22:23:45.63 & 1.4888 & \\
 1.2 & 11:49:35.86 & 22:23:50.78 & 1.4888 & \\
 1.3 & 11:49:36.82 & 22:24:08.73 & 1.4888 & \\
 2.1 & 11:49:36.58 & 22:23:23.06 & 1.894 & \\
 2.2 & 11:49:37.46 & 22:23:32.94 & 1.894 & \\
 2.3 & 11:49:37.58 & 22:23:34.37 & 1.894 & \\
 3.1 & 11:49:33.78 & 22:23:59.42 & 3.128 & \\
 3.2 & 11:49:34.25 & 22:24:11.07 & 3.128 & \\
 3.3 & 11:49:36.31 & 22:24:25.86 & 3.128 & \\
 4.1 & 11:49:34.32 & 22:23:48.57 & 2.95 & \\
 4.2 & 11:49:34.66 & 22:24:02.62 & 2.95 & \\
 4.3 & 11:49:37.01 & 22:24:22.03 & 2.95 & \\
 5.1 & 11:49:35.94 & 22:23:35.02 & 2.79 & \\
 5.2 & 11:49:36.27 & 22:23:37.77 & 2.79 & \\
 5.3 & 11:49:37.91 & 22:24:12.74 & 2.79 & \\
 6.1 & 11:49:35.93 & 22:23:33.16 & [2.0-3.0] & 2.72$^{+0.08}_{-0.06}$\\
 6.2 & 11:49:36.43 & 22:23:37.89 &  & \\
 6.3 & 11:49:37.93 & 22:24:09.02 &  & \\
 7.1 & 11:49:35.75 & 22:23:28.80 & [2.0-3.0] & 2.63$^{+0.09}_{-0.06}$\\
 7.2 & 11:49:36.81 & 22:23:39.37 &  & \\
 7.3 & 11:49:37.82 & 22:24:04.47 &  & \\
 8.1 & 11:49:35.64 & 22:23:39.66 & [2.0-3.0] & 2.97$^{+0.03}_{-0.03}$\\
 8.2 & 11:49:35.95 & 22:23:42.20 &  & \\
 8.3 & 11:49:37.70 & 22:24:16.99 &  & \\
 9.1 & 11:49:37.24 & 22:25:34.40 & 0.981 & \\
 9.2 & 11:49:36.93 & 22:25:37.98 & 0.981 & \\
 9.3 & 11:49:36.78 & 22:25:38.00 & 0.981 & \\
 9.4 & 11:49:36.88 & 22:25:35.07 & 0.981 & \\
 10.1 & 11:49:37.07 & 22:25:31.83 & [1.0-1.5] & 1.31$^{+0.09}_{-0.06}$\\
 10.2 & 11:49:36.87 & 22:25:32.26 &  & \\
 10.3 & 11:49:36.53 & 22:25:35.80 &  & \\
 13.1 & 11:49:36.89 & 22:23:52.03 & [1.0-1.5] & 1.28$^{+0.02}_{-0.01}$\\
 13.2 & 11:49:36.68 & 22:23:47.96 &  & \\
 13.3 & 11:49:36.01 & 22:23:37.89 &  & \\
 14.1 & 11:49:34.00 & 22:24:12.61 & [2.5-4.0] & 2.55$^{+1.07}_{-0.06}$\\
 14.2 & 11:49:33.80 & 22:24:09.45 &  & \\
 15.1 & 11:49:38.21 & 22:23:15.70 & [2.0-8.0] & 3.38$^{+0.15}_{-0.14}$\\
 15.2 & 11:49:38.48 & 22:23:19.48 &  & \\
 15.3 & 11:49:37.50 & 22:23:07.26 &  & \\
 16.1 & 11:49:38.33 & 22:23:15.58 & [1.0-6.0] & 4.83$^{+1.94}_{-1.44}$\\
 16.2 & 11:49:38.37 & 22:23:16.18 &  & \\
 17.1 & 11:49:38.39 & 22:23:14.04 & [1.0-7.0] & 4.23$^{+0.33}_{-0.29}$\\
 17.2 & 11:49:38.70 & 22:23:18.45 &  & \\
 17.3 & 11:49:37.58 & 22:23:04.14 &  & \\
 18.1 & 11:49:38.30 & 22:23:11.98 & [1.0-8.0] & 5.04$^{+0.64}_{-0.56}$\\
 18.2 & 11:49:38.90 & 22:23:20.61 &  & \\
 18.3 & 11:49:37.61 & 22:23:03.55 &  & \\
 21.1 & 11:49:34.28 & 22:24:46.33 & [2.0-3.0] & 2.57$^{+0.06}_{-0.07}$\\
 21.2 & 11:49:34.45 & 22:24:47.10 &  & \\
 21.3 & 11:49:34.81 & 22:24:45.67 &  & \\
 22.1 & 11:49:36.96 & 22:23:34.44 & 3.216 & \\
 22.2 & 11:49:38.17 & 22:24:00.84 & 3.216 & \\
 22.3 & 11:49:36.04 & 22:23:24.54 & 3.216 & \\
 \multicolumn{5}{c}{...} \\
\hline
\end{tabular}
\end{table*}
\begin{table*}
\centering
\begin{tabular}{lllll}
\hline
ID & $\alpha$ & $\delta$ & $z_{\rm prior}$ & $z_{\rm model}$ \\
   & [deg] & [deg] & & \\
\hline 
\multicolumn{5}{c}{...} \\
 26.1 & 11:49:37.14 & 22:25:33.52 & [0.6-1.5] & 0.97$^{+0.07}_{-0.07}$\\
 26.2 & 11:49:36.87 & 22:25:33.88 &  & \\
 26.3 & 11:49:36.66 & 22:25:36.97 &  & \\
 29.1 & 11:49:37.92 & 22:23:20.60 & [2.0-4.0] & 2.74$^{+0.08}_{-0.15}$\\
 29.2 & 11:49:38.18 & 22:23:25.46 &  & \\
 29.3 & 11:49:37.08 & 22:23:12.13 &  & \\
 31.1 & 11:49:36.52 & 22:23:48.29 & [2.0-3.0] & 2.66$^{+0.11}_{-0.05}$\\
 31.2 & 11:49:34.87 & 22:23:30.60 &  & \\
 31.3 & 11:49:37.35 & 22:24:08.78 &  & \\
 34.1 & 11:49:37.97 & 22:23:17.22 & [2.0-5.0] & 3.41$^{+0.19}_{-0.15}$\\
 34.2 & 11:49:38.49 & 22:23:26.24 &  & \\
 34.3 & 11:49:37.24 & 22:23:09.71 &  & \\
\hline
\end{tabular}
\caption{Multiple images used in the MACS1149 models. The results on the redshifts estimation are the ones of Model \texttt{III}.}
\label{tab:m1149mul}
\end{table*}

\begin{table*}
\centering
\begin{tabular}{lllllllllll}
\hline
Potential & $\Delta\alpha$ & $\Delta\delta$ & $e$ & $\theta$ & r$_{\rm core}$ & r$_{\rm cut}$ & $\sigma_{0}$ & $\gamma$ \\
   & [arcsec] & [arcsec] & & [deg] & kpc & kpc & km/s \\
\hline 
\multicolumn{8}{l}{Model \texttt{I}} \\
GGL & $[-13.6]$ & $[137.2]$ & $[0.17]$ & $[120.0]$ & $[0]$ & $[50]$ & $190^{+5}_{-2}$ \\
\hline
   \hline 
\multicolumn{8}{l}{Model \texttt{II}} \\
GGL & $[-13.6]$ & $[137.2]$ & $[0.17]$ & $[120.0]$ & $[0]$ & $[50]$ & $193^{+5}_{-5}$ \\
Ext Shear &  &  &  & $ 7.2^{+ 10.4}_{ -5.0} $ &  &  & & 0.13$^{+0.08}_{-0.06}$ \\
\hline
\hline 
\multicolumn{8}{l}{Model \texttt{III}} \\
GGL & $[-13.6]$ & $[137.2]$ & $[0.17]$ & $[120.0]$ & $[0]$ & $[50]$ & $174^{+28}_{-2}$ \\
DM1 & $ -3.2^{+  0.3}_{ -0.3}$ & $  1.4^{+  0.2}_{ -0.2}$ & $ 0.56^{+ 0.01}_{-0.01}$ & $ 40.0^{+  0.5}_{ -0.3}$ & $92^{+2}_{-3}$ & $[1000]$ & $1015^{+5}_{-11}$ \\
DM2 & $-23.7^{+  0.9}_{ -0.5}$ & $-28.0^{+  1.0}_{ -1.2}$ & $ 0.17^{+ 0.06}_{-0.04}$ & $128.6^{+  6.2}_{ -7.8}$ & $163^{+15}_{-25}$ & $[1000]$ & $124^{+32}_{-34}$ \\
DM3 & $-43.0^{+  0.4}_{ -1.0}$ & $-53.0^{+  0.4}_{ -0.4}$ & $ 0.64^{+ 0.08}_{-0.03}$ & $ 30.1^{+  3.8}_{ -6.4}$ & $44^{+15}_{-15}$ & $[1000]$ & $403^{+27}_{-26}$ \\
DM4 & $ 18.9^{+  0.5}_{ -0.3}$ & $ 47.2^{+  1.5}_{ -0.7}$ & $ 0.65^{+ 0.09}_{-0.05}$ & $124.9^{+  8.4}_{ -9.2}$ & $142^{+9}_{-9}$ & $[1000]$ & $482^{+40}_{-21}$ \\
DM5 & $-17.4^{+  0.4}_{ -0.4}$ & $101.0^{+  0.3}_{ -0.3}$ & $ 0.53^{+ 0.08}_{-0.03}$ & $129.5^{+  5.1}_{ -9.4}$ & $9^{+4}_{-1}$ & $[1000]$ & $354^{+29}_{-11}$ \\
BCG  & $[  0.0]$ & $[  0.0]$ & $[0.20]$ & $[ 34.0]$ & $36^{+3}_{-3}$ & $118^{+21}_{-23}$ & $256^{+21}_{-24}$ \\
GAL1 & $[  3.2]$ & $[-11.1]$ & $ 0.56^{+ 0.03}_{-0.03}$ & $ 45.9^{+ 10.9}_{ -6.3}$ & $[0]$ & $68^{+2}_{-1}$ & $208^{+10}_{-9}$ \\
L$^{*}$ galaxy &  & & & & $[0.15]$ & $44^{+3}_{-3}$ & $198^{+2}_{-2}$\\
\hline
\hline 
\multicolumn{8}{l}{Model \texttt{IV}} \\
DM1 & $ -4.1^{+  0.1}_{ -0.6}$ & $  1.3^{+  0.1}_{ -1.1}$ & $ 0.60^{+ 0.02}_{-0.01}$ & $ 28.8^{+  0.6}_{ -6.6}$ & $99^{+2}_{-20}$ & $[1000]$ & $899^{+7}_{-62}$ \\
DM2 & $-25.3^{+  0.2}_{ -0.2}$ & $-32.8^{+  0.8}_{ -0.3}$ & $ 0.70^{+ 0.12}_{-0.01}$ & $ 49.6^{+  0.8}_{ -8.9}$ & $66^{+7}_{-33}$ & $[1000]$ & $442^{+12}_{-16}$ \\
DM3 &  $-48.3^{+  0.6}_{ -3.7}$ & $-49.6^{+  0.3}_{ -1.1}$ & $ 0.39^{+ 0.02}_{-0.04}$ & $175.8^{+  6.6}_{-36.3}$ & $221^{+10}_{-7}$ & $[1000]$ & $481^{+41}_{-11}$ \\
DM4 & $ 23.3^{+  0.1}_{ -1.1}$ & $ 47.2^{+  1.3}_{ -0.2}$ & $ 0.26^{+ 0.01}_{-0.11}$ & $103.1^{+  2.3}_{-10.9}$ & $76^{+2}_{-23}$ & $[1000]$ & $584^{+31}_{-11}$ \\
DM5 & $-16.3^{+  0.1}_{ -0.2}$ & $100.3^{+  0.0}_{ -0.0}$ & $ 0.24^{+ 0.01}_{-0.09}$ & $130.1^{+  3.6}_{ -9.0}$ & $2^{+0}_{-1}$ & $[1000]$ & $444^{+6}_{-3}$ \\
BCG & $[  0.0]$ & $[  0.0]$ & $[0.20]$ & $[ 34.0]$ & $34^{+0}_{-5}$ & $258^{+1}_{-12}$ & $373^{+28}_{-7}$ \\
GAL1 &  $[  3.2]$ & $[-11.1]$ & $ 0.02^{+ 0.01}_{-0.09}$ & $ 94.3^{+  6.7}_{ -2.1}$ & $[0]$ & $44^{+1}_{-3}$ & $171^{+2}_{-10}$ \\
L$^{*}$ galaxy &  & & & & $[0.15]$ & $67^{+5}_{1}$ & $143^{+1}_{-18}$\\
\hline
\end{tabular}
\caption{Table of the parameters of the models of MACS1149.}
\label{tab:m1149params}
\end{table*}

\begin{table*}
\centering
\begin{tabular}{lllll}
\hline
ID & $\alpha$ & $\delta$ & $z_{\rm prior}$ & $z_{\rm model}$ \\
   & [deg] & [deg] & & \\
\hline 
 A.1 & 03:29:40.74 & -02:13:17.90 & 1.112 & \\
 A.2 & 03:29:40.85 & -02:13:17.19 & 1.112 & \\
 A.3 & 03:29:40.87 & -02:13:17.44 & 1.112 & \\
 B.1 & 03:29:40.75 & -02:13:18.15 & 1.112 & \\
 B.2 & 03:29:40.83 & -02:13:17.17 & 1.112 & \\
 B.3 & 03:29:40.88 & -02:13:17.93 & 1.112 & \\
 B.4 & 03:29:40.85 & -02:13:18.54 & 1.112 & \\
 1.1 & 03:29:40.17 & -02:11:45.71 & 6.18 & \\
 1.2 & 03:29:40.07 & -02:11:51.71 & 6.18 & \\
 1.3 & 03:29:41.24 & -02:12:04.66 & 6.18 & \\
 1.4 & 03:29:43.16 & -02:11:17.36 & 6.18 & \\
 2.1 & 03:29:41.03 & -02:11:29.06 & 2.14 & \\
 2.2 & 03:29:39.62 & -02:12:00.66 & 2.14 & \\
 2.3 & 03:29:42.17 & -02:11:25.61 & 2.14 & \\
 2.4 & 03:29:42.33 & -02:11:54.46 & 2.14 & \\
 3.1 & 03:29:40.18 & -02:11:26.56 & [1.0-5.0] & 2.58$^{+0.05}_{-0.05}$\\
 3.2 & 03:29:39.06 & -02:11:49.91 &  & \\
 3.3 & 03:29:41.26 & -02:11:15.16 &  & \\
\hline
\end{tabular}
\caption{Multiple images used for the models of MACS0329. The results on the redshifts estimation are the ones of Model \texttt{III}.}
\label{tab:m0329mul}
\end{table*}

\begin{table*}
\centering
\begin{tabular}{llllllllllll}
\hline 
Potential & $\Delta\alpha$ & $\Delta\delta$ & $e$ & $\theta$ & r$_{\rm core}$ & r$_{\rm cut}$ & $\sigma_{0}$ & $\gamma$ \\
   & [arcsec] & [arcsec] & & [deg] & kpc & kpc & km/s \\
   \hline 
\multicolumn{8}{l}{Model \texttt{I}} \\
GGL & $[ 11.2]$ & $[-91.4]$ & $[0.40]$ & $[ 74.0]$ & $[0]$ & $[50]$ & $209^{+1}_{-1}$ \\
\hline
   \hline 
\multicolumn{8}{l}{Model \texttt{II}} \\
GGL & $[ 11.2]$ & $[-91.4]$ & $[0.40]$ & $[ 74.0]$ & $[0]$ & $[50]$ & $196^{+2}_{-6}$ \\
Ext Shear &  &  &  & $  4.0^{+ 11.3}_{ -2.1}$ &  &  & & 0.17$^{+0.08}_{-0.04}$ \\
\hline
\hline
\multicolumn{8}{l}{Model \texttt{III}} \\
GGL & $[ 11.2]$ & $[-91.4]$ & $[0.40]$ & $[ 74.0]$ & $[0]$ & $[50]$ & $188^{+4}_{-9}$ \\
DM1 & $ -1.4^{+  0.4}_{ -0.4}$ & $ -0.7^{+  0.4}_{ -0.2}$ & $ 0.25^{+ 0.03}_{-0.01}$ & $ 70.1^{+  1.8}_{ -2.7}$ & $58^{+19}_{-3}$ & $[1000]$ & $959^{+28}_{-19}$ \\
DM2 & $ 39.4^{+  5.0}_{ -1.8}$ & $ 22.1^{+  7.4}_{ -2.0}$ & $ 0.46^{+ 0.17}_{-0.10}$ & $ 98.0^{+  6.5}_{ -6.0}$ & $119^{+0}_{-19}$ & $[1000]$ & $877^{+40}_{-38}$ \\
BCG & $[ -0.0]$ & $[  0.0]$ & $[0.19]$ & $[-73.6]$ & $[0]$ & $[98]$ & $208^{+11}_{-193}$ \\
L$^{*}$ galaxy &  & & & & $[0.15]$ & $[45]$ & $155^{+6}_{-6}$\\
\hline
\hline 
\multicolumn{8}{l}{Model \texttt{IV}} \\
DM1 & $ -1.4^{+  0.4}_{ -0.4}$ & $ -0.7^{+  0.4}_{ -0.3}$ & $ 0.24^{+ 0.03}_{-0.02}$ & $ 69.4^{+  1.8}_{ -3.2}$ & $59^{+19}_{-4}$ & $[1000]$ & $984^{+24}_{-15}$ \\
DM2 & $ 38.2^{+  5.4}_{ -2.5}$ & $ 20.7^{+  8.6}_{ -1.8}$ & $ 0.48^{+ 0.16}_{-0.11}$ & $ 98.3^{+  2.2}_{-11.2}$ & $114^{+0}_{-12}$ & $[1000]$ & $833^{+37}_{-36}$ \\
BCG & $[ -0.0]$ & $[  0.0]$ & $[0.19]$ & $[-73.6]$ & $[0]$ & $[98]$ & $75^{+14}_{-242}$ \\
L$^{*}$ galaxy &  & & & & $[0.15]$ & $[45]$ & $157^{+5}_{-7}$\\
\hline
\end{tabular}
\caption{Parameters for the models of MACS0329}
\label{tab:m0329params}
\end{table*}


\bsp	
\label{lastpage}
\end{document}